\documentclass[acmsmall,nonacm]{acmart}

\renewcommand\footnotetextcopyrightpermission[1]{} 
\AtBeginDocument{%
  }

\setcopyright{acmlicensed}
\copyrightyear{2025}
\acmYear{2025}
\acmDOI{XXXXXXX.XXXXXXX}

\acmConference[Some]{Some}{2025}{Place}
\acmISBN{978-1-4503-XXXX-X/18/06}

\usepackage{paralist}
\usepackage{dcolumn}
\usepackage{enumitem}
\usepackage{xcolor}
\usepackage{graphicx}
\usepackage{soul}
\usepackage{fixme}
\definecolor{mURLs}{rgb}{0.13,0.26,0.39}	
\usepackage{hyperref}
\newcommand{\supp}{Appendix} 

\usepackage{tcolorbox}
\tcbuselibrary{skins,breakable,listings}

\tcbset{
  llmprompt/.style={
    enhanced,
    width=\linewidth, 
    colback=white,
    colframe=gray!35,
    coltitle=black,
    colbacktitle=gray!10,
    fonttitle=\bfseries\small\sffamily,
    arc=1.5mm,
    boxrule=1pt,
    top=1mm, bottom=1mm, left=1mm, right=1mm,
    boxsep=4pt,
    before skip=10pt, after skip=10pt,
    breakable,
    listing only,
    listing options={
      basicstyle=\ttfamily\small,
      breaklines=true,
      breakatwhitespace=false,
      columns=flexible,
      keepspaces=true,
      showstringspaces=false
    },
  }
}

\begin{document}

\title{Cognitive Spillover in Human-AI Teams}

\author{Christoph Riedl}
\authornote{Corresponding author.}
\affiliation{%
  \institution{Northeastern University}
  \city{Boston}
  \country{USA}}
\email{c.riedl@northeastern.edu}
\orcid{0000-0002-3807-6364}

\author{Saiph Savage}
\affiliation{%
  \institution{Northeastern University}
  \city{Boston}
  \state{MA}
  \country{USA}}
\email{s.savage@northeastern.edu}

\author{Josie Zvelebilova}
\affiliation{%
  \institution{Northeastern University}
  \city{Boston}
  \state{MA}
  \country{USA}}
\email{zvelebilova.j@northeastern.edu}

\renewcommand{\shortauthors}{Riedl et al.}

\begin{abstract}
AI is not only a neutral tool in team settings; it influence the social and cognitive fabric of collaboration. Across two randomized experiments, we demonstrate that AI exposure produces causal spillover into human-human interaction---affecting shared language, collective attention, shared mental models, and social cohesion. These spillover effects occur robustly across settings, modalities, tasks, and AI qualities, suggesting that mere exposure to AI drives the influence. AI functions as an implicit ``social forcefield,'' influencing not only how people speak, but also how they think, what they attend to, and how they relate to each other. We argue for shifting the design paradigm from optimizing ``AI as a tool'' to understanding AI as a socially influential actor whose effects extend beyond the human-AI interface.
\end{abstract}



\received{02 July 2024}
\received[revised]{Day Month 2024}
\received[accepted]{Day Month 2024}

\maketitle

\clearpage 
\section{Introduction}
Artificial Intelligence (AI) is increasingly pervasive in modern workplaces \cite{brynjolfsson2025generative,agarwal2023combining} and projected to have wide ranging impact on the future of work \cite{bcg2024aiadoption,aiindex2025}. 
Yet there is also growing unease about indirect consequences \cite{kennedy2025pew,chayka2025ai} and heightened regulatory interest in systemic AI risk \cite{bengio2024managing}.
In particular, AI use does not occur in isolation as workers are embedded in teams where they are engaged in social interactions with human coworkers. The key thesis of this paper is that AI systems do not just passively assist users; they influence the social and cognitive fabric of teamwork beyond their direct interactions with the AI. By shifting the analytical lens from ``AI as tool'' to ``AI as social forcefield,'' we show how AI influence can spill over to human-human interaction beyond AI's immediate functional role.

We draw on prior literature on distributed cognition \cite{hutchins1995cognition,hollan2000distributed} to articulate how AI influences cognitive alignment in human–AI teams \cite{pickering2004toward,garrod2009joint}. A core insight from this literature is that alignment operates simultaneously across multiple coupled channels: linguistic, attentional, representational, and affective \cite{garrod2009joint,clark1996using}. Shared language enables groups to coordinate how they understand situations in real time \cite{clark1996using,sievers2024consensus,tausczik2010liwc}; collective attention orients groups toward common objects and topics \cite{kommol2023components,woolley2023collective,riedl2017teams,ocasio1997towards}; shared mental models align interpretive frameworks and behavioral expectations \cite{fusaroli2012coming,hollan2000distributed,dechurch2010cognitive}; and the affective sense that ``we understand each other,'' shapes group identity and social cohesion \cite{shteynberg2015shared,tomasello2009cultural,hutchins1995cognition}. While alignment at one level can promote alignment at others, these channels operate in parallel rather than in strict causal sequence \cite{pickering2004toward}. These constructs span the depth of collaborative dynamics, from automatic (linguistic) to deliberative (cognitive) to affective (social), capturing how the introduction of AI can affect coupled alignment processes within human teams.

A major challenge in measuring AI's influence in team settings is distinguishing between direct human-AI \textit{alignment} and \textit{spillover} into subsequent human-human interaction, while controlling for the natural alignment resulting from shared task history alone. It is well known that people align with their \textit{current} conversation partner's language (e.g., word choice, syntax, style) \cite{wachsmuth2013alignment,brennan1996conceptual,garrod1987saying}, even when that partner is an AI \cite{branigan2011role,Hu2014Entrainment}. It is also known that interaction dynamics themselves naturally produce shared language even without explicit interpersonal influence \cite{garrod1987saying,brennan1996conceptual,clark1986referring}. To claim that AI actively shapes team cognition, one must rigorously disentangle causal AI influence from the natural baseline of task-driven vocabulary alignment and look beyond the (dyadic) influence during AI use itself.

Current HCI literature has not yet resolved this challenge. While recent work has made significant advances in understanding human-AI interaction \cite{rezwana2023designing,riedl2025quantifying,inkpen2023advancing,xiao2020tell,bansal2019beyond, branigan2011role, cowan2015voice, wudarczyk2021robots}, prior work focuses largely on effects within the human-AI dyad during active use \cite[e.g.,][]{Zajac2025,Vasconcelos2025,Ma2025}, without testing whether such effects produce spillover into human-human dynamics. Team-focused research reports mixed results---AI is sometimes helpful, sometimes harmful, often unpredictably variable \cite[e.g.,][]{autor2014polanyi, fugener2021will, dell2023navigating}---in part because it typically examines isolated outcomes like performance \cite{wang2022documentation} rather than the multi-channel alignment mechanisms that coordinate teamwork \cite{schelble2022let}. And studies of subtle AI influences---such as performance boosts from perceived AI support \cite{kosch2023placebo} or perceived ownership in AI-assisted writing \cite{draxler2024ai}---focus on effects during \textit{active} AI engagement \cite{chaves2022chatbots}, leaving open whether AI shapes social interaction among team members beyond their direct interaction with the AI. We address these gaps by testing for AI spillover across multiple parallel channels of socio-cognitive alignment---linguistic, attentional, cognitive, and relational---and tracing effects beyond active AI engagement into subsequent human-human interaction.

In this paper, we provide evidence of causal spillover in two randomized controlled studies while carefully controlling for shared-task baselines.
\textbf{Study 1} focuses specifically on isolating causal spillover of AI language to subsequent human-human interaction. Participants used ChatGPT 4o to respond to customer service complaints. By manipulating the AI's system prompt (empathic vs.~formal), we exposed users to systematic linguistic variations while holding the underlying task and interaction dynamic constant. We then analyzed users' subsequent language use during a face-to-face debriefing interview with a human partner. This design directly tests whether subtle differences in the AI assistant's task framing and language leave a linguistic and cognitive ``fingerprint'' that persists in subsequent human–human interaction beyond the shared task vocabulary.
\textbf{Study 2} expands this to a dynamic, ecologically complex team setting. Teams of 3--4 individuals performed a complex problem-solving task with a voice-based AI assistant. We manipulated AI attributes (helpfulness and voice anthropomorphism) and measured the effects on shared language, collective attention, shared mental models, and social cohesion.

This research answers three high-level questions:
\begin{enumerate}[label=RQ\arabic*]
    \item \textbf{Causal Spillover:} Does exposure to AI causally affect socio-cognitive alignment of human teams---linguistic, cognitive, and social---beyond the direct interaction with the AI and beyond natural alignment resulting from shared task alone?
    \item \textbf{Robustness:} Are these effects robust across different AI modalities (text vs.~voice), attributes (anthropomorphism and helpfulness), and phases of activity (AI-present vs.~AI-absent)?
    \item \textbf{Conscious vs.~Implicit Influence:} To what extent is this socio-cognitive alignment dependent on positive user appraisal (e.g., trust, perceived intelligence, perceived team membership), or does AI affect team dynamics through an implicit, automatic process regardless of user sentiment?
\end{enumerate}

In summary, we make the following contributions to the understanding of how AI influences socio-cognitive alignment in human-AI teams across multiple parallel channels. 
\begin{compactenum}
    \item \textbf{Empirical Evidence of Spillover:} We provide causal evidence that AI influence spills over into human-human interaction. By rigorously controlling for shared task vocabulary, we demonstrate that this effect is not merely dyadic entrainment, but a detectable ``residue'' in human-human communication. 
    \item \textbf{AI Influence on Multiple Channels:} We show that AI-induced spillover is not confined to surface-level linguistic mimicry but manifests across multiple parallel channels of socio-cognitive alignment---from automatic linguistic coordination to deliberative shared mental models to affective group cohesion. This breadth is nontrivial: spillover could plausibly have been absent, weak, or limited to a single channel \cite[e.g.,][]{pickering2004interactive,danescu2012echoes,glikson2020human,danescu2012echoes,gray2012feeling,reeves1996media}.
    \item \textbf{Generalizability Across Contexts:} By triangulating evidence across two heterogeneous studies (individual vs.~team, text vs.~voice, customer service vs.~problem-solving task, high- vs.~low-capacity agents), we show that these alignment effects are robust. Notably, they are detectable even when the AI is no longer present, has limited capacity, or when users report low trust, suggesting an implicit cognitive effect distinct from conscious reliance. The convergence of effects across different contexts provides stronger evidence of generalizable predictive validity than any single test \cite{campbell1959convergent,cook2002experimental,lawlor2016triangulation}.
    \item \textbf{Reconceptualization of AI in Teams:} We argue for shifting the design paradigm from optimizing ``AI as a tool'' to understanding ``AI as a social forcefield.'' The spillover effects we document emerge readily and affect different parallel channels of alignment, suggesting that AI's influence on team cognition is broader and less contained than current design frameworks assume.
\end{compactenum}

\section{Conceptual Framework for Socio-Cognitive Alignment in Human–AI Teams}
Effective collaboration in teams---whether human-only, human-AI, or AI-AI---requires that members represent the task and each other in mutually compatible ways \cite{hutchins1995cognition,riedl2025quantifying,riedl2026emergent}. 
This representational alignment enables shared reference (``what are we talking about?''), coordinated attention (``what matters right now?''), prediction of others' behavior (``what will they do next?''), and effective coordination (``can I rely on them to interpret this the same way?''). Thus, alignment is not a single construct but a multi-channel process, observable at different levels of interaction \citep{hutchins1995cognition,clark1996using,resnick1991perspectives}. We theorize that AI assistants---and especially AI language models---are not passive tools; instead, they mediate alignment by introducing linguistic and conceptual anchors that can persist beyond the direct (dyadic) interaction with the AI itself and influence the cognitive and social fabric of teamwork on different levels. Thus, our work extends insights on how  digital artifacts shape coordination \cite{hollan2000distributed,hutchins1995cognition} to the social dynamics of human–AI teams.\footnote{When we speak of human-AI teams we mean multiple humans not a single human-AI dyad.}

\begin{figure}[hbtp!]
  \centering
  \includegraphics[width=0.75\textwidth]{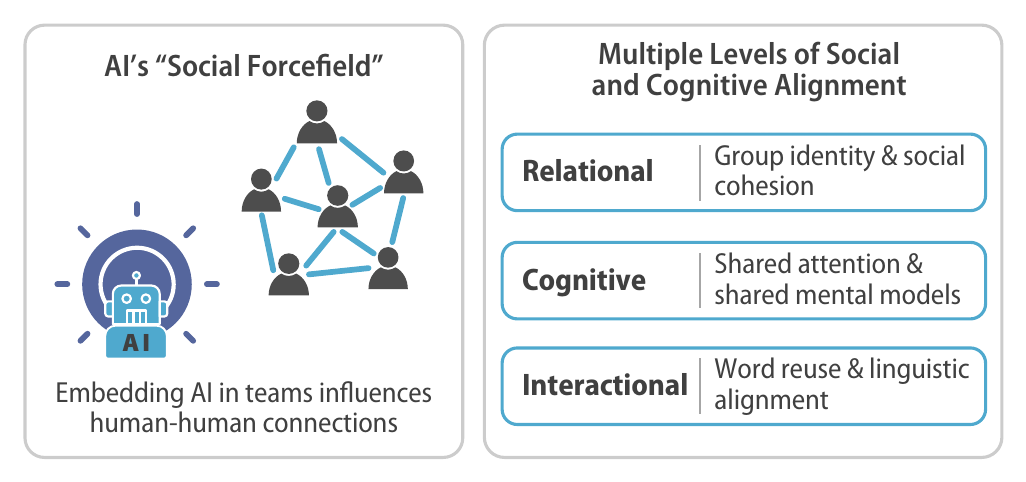}
  \caption{
  Conceptual framework of AI spillover on different levels of cognitive and social alignment.
  (Left) AI exerts a ``social forcefield,'' with its influence spilling over to human-human interactions.
  (Right) Three parallel channels: linguistic, cognitive, and social.}
  \label{fig:framework}
  \Description{Conceptual framework of AI spillover on different levels of cognitive and social alignment.}
\end{figure}

Grounded in prior work on distributed and team cognition \cite{hutchins1995cognition,clark1996using,resnick1991perspectives,cooke2000measuring}, we adopt a literature-based framework that points to alignment processes happening on multiple levels in parallel (Figure \ref{fig:framework}). We then operationalize each level using established measures and test whether AI affects outcomes on that level. This literature-based framework serves as an analytical lens to structure our inquiry and inform our measures. 


\subsection{Shared Language}
Shared language reflects cognitive coupling that allows individuals to align their representations as they adapt to each other's speech in real time. Lexical alignment occurs when conversational partners converge on similar words or phrases during interaction \cite{wachsmuth2013alignment}. Over time, these repetitions form ``conceptual pacts,'' which are agreements on how to refer to things within a group \cite{brennan1996conceptual}. Alignment on shared language reduce cognitive effort; without shared language, communication would be more difficult and increase processing demands \cite{metzing2003conceptual}. Shared language therefore helps groups coordinate attention, conserve mental resources, and collaborate more efficiently \cite{kalocinski2018interactive,hutchins1995cognition,ocasio1997towards}. 

Causally attributing spillover to AI influence requires paying careful attention to what the baseline should be \cite{pickering2004interactive,clark1986referring}. Convergence on shared task vocabulary naturally arises from task structure \cite{garrod1987saying}, common ground \cite{clark1991grounding}, and coordination history \cite{brennan1996conceptual}. A ``no influence'' baseline would therefore be too lenient as it risks attributing naturally occurring alignment to AI-induced spillover. To demonstrate causal AI spillover, we need to establish effects above and beyond this baseline level of alignment (from task structure, common ground, coordination history, etc). Hence, our analyses operationalize AI spillover as the reuse of AI-specific lexical items in subsequent human-human interaction above baseline alignment on shared task vocabulary (see Methods for details). Whereas traditional forced-naming tasks may overestimate effects by capturing direct lexical substitution and anchoring \cite{brandstetter2017persistent}, our approach measures whether AI exposure shapes the language people naturally use, even when no explicit labeling decision is required. For example, in our second study, teams are not prompted to talk about the object mentioned by the AI but may choose to do so at their own volition.

\subsection{Collective Attention and Shared Mental Models}
Human attention is the ability to actively process specific information while tuning out other, less relevant details \cite{bundesen1990theory}. Attention is limited in both capacity and duration \cite{duncan2016visual}, making its effective coordination essential for high performance \cite{duncan2013structure}. In group contexts, individuals align their attention with others, including what they perceive, remember, and prioritize \cite{kampis2020altercentric}. Groups achieve high collective attention when members co-attend to the same object or event, creating a shared psychological experience that shapes motivation, judgment, and behavior \cite{shteynberg2015shared,woolley2023collective}. Collective attention thus enables teams to coordinate activities and sustain joint action \cite{mayo2021variance}. 
Prior work establishes that shared language plays a central role in shaping collective attention because it links cognitive processes to shared meaning, aligning beliefs and neural activity across team members \cite{nagai2003constructive,kaplan2006challenges}. By shaping how tasks are described and discussed, language influences collective attention through its alignment of intentions, plans, and strategies \cite{hawkins2023partners,ocasio1997towards}. We thus quantify collective attention via time-resolved analyses of conversational content---when does the team talk about what?---using topic timing/synchrony derived from transcripts \cite{weng2012competition,woolley2023collective}.

Together with collective attention, shared mental models are the collective cognitive structures that teams develop to interpret and predict their environment \cite{dechurch2010cognitive}. They capture a group's shared understanding of ``what is happening, what is likely to happen next, and why it is happening'' \cite[p.~879]{mohammed2010metaphor}. Such shared mental models improve motivation, coordination, and performance, particularly when tasks require mutual dependence on diverse skills and expertise \cite{lewis2003measuring}. 
Building on this literature, we operationalize the alignment of shared mental models using a validated and widely used shared mental models survey instrument (Study 2) \cite{johnson2007measuring}. 
We also capture alignment of mental models indirectly, drawing on collective framing \cite{mathieu2022indexing} so that when subjects utilize team-framed language to describe task execution, they are verbalizing a shared mental model (Study 1).

\subsection{Group Identity and Social Cohesion}
Social cohesion and group identity represent the affective dimension of collaborative alignment \cite{pickering2004interactive,clark1986referring}. When groups align across channels---linguistic, attentional, and representational---they experience cognitive fluency and affective coherence: the feeling that ``we understand each other'' \cite{hutchins1995cognition,echterhoff2009shared}. Convergence itself signals affiliation with the group \cite{coupland1991accommodation}. Shared language, coordinated attention, and aligned mental models thus co-occur with relational bonds and a sense of ``we-ness'' \cite{tomasello2009cultural,shteynberg2015shared,hinds2014language}.
Crucially, social cohesion depends on who is considered part of the group. Teams differentiate between in-group and out-group members, and this boundary shapes how much alignment occurs \cite{brennan2009partner,liberman2018children}. When an agent---such as an AI system---is treated as a teammate, it may facilitate greater alignment and integration of perspectives \cite{wang2021towards}, but if it is seen as an outsider, alignment may be weaker or even resisted \cite{xie2023two,mahmud2023study}. Whether AI is considered an insider or outsider is critical to understand its impact on the overall social fabric of the human team members \cite{groom2011responses,nass2000machines}.

Our operationalization of social cohesion builds on the idea of the dual role of language: language is not only a constructor of group identity but also serves as an index of it \cite{echterhoff2009shared,hutchins1995cognition,clark1996using}. In particular, language can signal collective identity through the use of function words that require mutual understanding, such as pronouns \cite{VanSwol2021,sievers2024consensus}. Groups with frequent pronoun use have a higher level of trust in their teammates and strong social cohesion \cite{tausczik2010liwc}. We consequently operationalize our measure of alignment on the relational layer through groups' use of pronouns \cite{tausczik2010liwc}. A substantial body of work has used function-word choices---particularly pronouns---as indicators of interpersonal alignment, collective identity, and cohesion \cite{tausczik2010liwc,gonzales2010language,pickering2014interactive,hutchins1995cognition,searle1995construction}. We supplement pronoun use as linguistic proxy with survey-based measures of group identity and cohesion to explore the extent to which AI instantiates these patterns.

In summary, we adopt a framework in which AI-induced spillover manifests across parallel channels of socio-cognitive alignment---linguistic, cognitive, and affective---spanning processes that range from automatic to deliberative to social. This breadth is theoretically meaningful: AI spillover could plausibly be confined to automatic linguistic processes alone, without affecting the full spectrum of collaborative cognition.

\section{Related Work}
Our research is most closely related to work on (a) human-AI teaming, and (b) alignment on shared language in human-AI teams.

\subsection{Human-AI Teaming}
The growing importance of teamwork \cite{wuchty2007increasing} has converged with rapid advances in artificial intelligence to fuel the development of collaborative human-AI teams \cite{malone2015handbook,oneill2020human}. The vision is that such teams can leverage the complementary strengths of humans and AI, thereby transcending the limitations of human-only groups \cite{brynjolfsson2018artificial,vaccaro2024combinations,riedl2025quantifying}.  

\subsubsection{From Task Performance to Cognitive Alignment}
Most research on human-AI interaction has emphasized performance outcomes, focusing on how AI systems automate processes, augment human decision-making, or generate novel solutions \cite[e.g.,][]{dell2023navigating,chen2024large,meincke2024prompting}. Experimental setups often compare human-only performance to human-AI dyad performance, with measures such as speed, accuracy, and output quality \cite{vaccaro2024combinations}. For example, studies in medical imaging show that AI can improve diagnostic accuracy when combined with radiologists’ judgments \cite[e.g.,][]{agarwal2023combining,Zajac2025}, and that AI collaboration can enhance decision-making performance \cite[e.g.,][]{fugener2022cognitive}. However, results are often inconsistent and true synergy can be difficult to achieve: some studies report performance gains, while others find disruption or reduced effectiveness \cite[e.g.,][]{autor2014polanyi,fugener2021will,dell2023navigating,vaccaro2024combinations}. This variability highlights the need to look beyond performance as a singular outcome and investigate the mechanisms that shape collaboration processes overall, such as collective attention, shared mental models, and social cohesion \cite{riedl2025quantifying,kelley2025personalized}. Several studies have started to shed light on different aspects of cognitive alignment \cite{bansal2019beyond,bansal2019updates,paleja2022aaai,alipour2021improving,bansal2019beyond,fugener2022cognitive}; most notably Schelble et al.~\cite{schelble2022let} studied team cognition, performance, and trust. Deeper understanding of how AI exposure affects alignment across different channels could help explain why performance benefits sometimes fail to materialize.

\subsubsection{Toward Multi-Member Dynamics}
Most studies of human-AI teaming remain focused on human-AI dyads \cite[e.g.,][]{bansal2019beyond,bansal2019updates,wang2019human,fugener2022cognitive,alipour2021improving}, leaving fundamental questions about how AI participation alters multi-member team dynamics \cite{dennis2023ai}. Collective cognition is not simply the sum of individual cognition but emerges from multi-agent coordination among human teammates and AI \cite{Demir2019The,Harrison2003TIME}. Prior research suggests these dynamics can be unpredictable: if one person adapts smoothly to the AI while another resists, overall team coordination may falter \cite{Balachandar2019Collaboration,Reichel2018Entrainment}. Similarly, in-group and out-group dynamics are inherently group phenomena that go beyond the human-AI dyad of direct interaction. Few studies have explored AI's effect on human-human interaction beyond the direct interaction with the AI itself.

\subsection{Alignment on Shared Language in Human-AI Interaction}
An emerging body of work explores how humans adapt linguistically to AI. \citet{spillner2021alignment} showed that users exhibit lexical alignment with chatbots in digital companion settings, while \citet{li2025co} provided a framework for bidirectional alignment, emphasizing both human-to-AI and AI-to-human processes. However, this work generally focuses on dyadic interactions, leaving open the question of how AI-generated language influences dynamics in larger human groups and beyond the direct interaction with (or presence of) the AI system.  \citet{guingrich2024ascribing} hypothesized that AI-generated language could spill over into human--human collaboration even after the AI is removed, but provided no empirical evidence. Our work directly investigates this proposition, showing how linguistic spillover from AI reshapes communication within human teams and leaves durable traces in group interaction. Brandstetter et al.~\cite{brandstetter2017persistent} provided insights on linguistic spillover from interaction with robots to subsequent word choices. We expand this work by focusing on word choices during naturalistic language production going beyond naming symbols in a forced-choice task. Furthermore, we introduce precise controls for baseline effects (such as lexical priming tendency and anchoring bias). That is, we capture representational spillover in new contexts, new utterances, new pragmatic roles, not just referential replacement.

When people talk, they co-construct shared terminology to maintain common ground \cite{branigan2011role,garrod1987saying}. With AI, the dynamics are more complicated and empirical evidence is mixed. Some studies suggest stronger alignment with AI than with human partners \cite{branigan2011role}---especially if they perceive the AI as less capable of adapting to them \cite{branigan2011role,Hu2014Entrainment,Weatherholtz2014Socially-mediated}---while others find the opposite \cite{wudarczyk2021robots,brandstetter2017persistent}. Still, most of this work focuses on dyads, leaving unanswered how alignment unfolds in mixed human-AI groups beyond the direct interaction with AI and its broader consequences beyond just language alignment. Our study addresses this gap by showing how spillover from AI can leave lasting linguistic and cognitive traces in human-human interactions.

\subsection{Summary of Research Gaps}
This review or related work highlights three primary gaps. First, existing studies of human-AI alignment are often limited to dyadic interactions (one-human-one-AI) and situations of direct AI interaction, leaving open questions about how AI spills over to multi-member teams and situations in which the AI may not be present. 
Second, research on human-AI collaboration has focused too narrowly on performance as task outcome, neglecting the cognitive and social processes that underpin teamwork, thus making it difficult to explain situations in which AI fails to deliver performance improvements. 
Third, foundational work in human communication shows that alignment and entrainment are central to cognitive alignment, suggesting that AI could significantly reshape alignment on different cognitive and social levels. Our work addresses all three gaps: we test for spillover from human-AI to human-human interaction, measure effects across multiple parallel channels of collaborative alignment, and trace these effects beyond the boundaries of active AI engagement.

\section{Study 1 - Customer Service Encounters}
To establish proof for the existence of causal linguistic spillover, we designed a preregistered\footnote{\url{https://osf.io/quj2d/?view_only=9299fa86594b487c9c356e1a32dbe6f7}} and IRB approved study. This study uses a minimal design with a state-of-the-art interactive chatbot. It directly tests whether subtle variations in an AI assistant's framing and language can leave a cognitive and linguistic ``fingerprint'' that is detectable when people later communicate with other humans---even when the AI is no longer present. The study design randomly manipulates the AI's system prompt---affecting subtle word choices while keeping the overall task and interaction dynamics constant---and measures how much subjects spontaneously reuse AI words in a face-to-face debriefing interview. This experimental manipulation of linguistic exposure thus isolates the causal pathway through which AI priming reshapes temporally disjoint human language and thought beyond the direct interaction with the AI itself. The only systematic difference between treatments is the AI's linguistic framing, which allows us to isolate spillover specific to the AI's style rather than generic entrainment based on common ground or shared tasks.

\subsection{Study Design and Task}
To achieve high ecological validity, we designed a study around a realistic task: AI assisting customer support agents who perform text-based chat support to resolve customer complaints and technical problems. AI is already widely used for such tasks \cite{bcg2024aiadoption}. One prominent field study in a Fortune 500 company showed that AI is highly effective supporting customer service agents, reporting an 14\% average increase in productivity \cite{brynjolfsson2025generative}. Our study had two phases. First, users interacted with a state-of-the-art chatbot (ChatGPT 4o) \cite{hurst2024gpt} to write answers to five customer service complaints across different domains (including technical trouble shooting, online shopping assistance, and reservation booking; one example service complaint was ``Every time I try to watch a movie, it buffers endlessly''). The AI was a fully functional ChatGPT 4o assistant: subjects could interact with it like with any chatbot system, write any prompt they liked, for as many rounds as they wanted. Subjects interacted with the AI on a study-provided laptop computer. Users were randomly assigned to one of two AI conditions. Through the system prompt we instructed ChatGPT to either suggest customer service responses that emphasized warmth and empathy or formal adherence to company guidelines (see full system prompts in \supp~\ref{appendix:study2}). The system prompt manipulation subtly changed which words subjects were exposed to by the AI. We term these \textit{exposure words} and they provide the key experimental manipulation and basis for our analysis. We recorded all user-written prompts and ChatGPT responses (i.e., exposure words), as well as the ultimate response to the customer complaint through a custom user interface implemented via Qualtrics. Each conversation started with ChatGPT offering a suggested draft response to the customer complaint. Subjects wrote on average one additional prompt per customer complaint to refine the initial draft response generated by the chatbot ($SD=1$). The typical interaction was two prompts long. We find no significant different in engagement between the two treatment conditions (two sample t-test, $t = 0.843$, $df = 92.988$, $p = 0.401$).

After the interaction with the AI, subjects participated in an in-person, face-to-face debriefing interview with an experimenter that lasted about five minutes (SD = 108 seconds). The debriefing interview questions were exclusively focused on the tasks---not the AI. To capture not only how the AI shapes how people talk, but also how they imagine acting in the future---their cognitive-behavioral approach to the task---questions were specifically designed to elicit subject's process, goal, or approach to the task. For example, one question from the interview guide was ``If you were working on a real customer service team, what kind of approach would you suggest the team agree on for answering complaints?'' A common follow-up question explicitly elicited subject's approach to the task and behavioral intention ``What's the first thing you usually focus on [when answering customer service complaints]?'' In this way, our measure of overlap captures not only linguistic spillover but also how the AI interaction influences how people think about the task itself, thus capturing important aspects of cognitive alignment through which groups sustain a common frame of reference for joint action \cite{shteynberg2015shared,clark1996using,clark1991grounding,tomasello2009cultural,hutchins1995cognition}.

Interviews were explicitly framed around team-level constructs, asking subjects to describe how a team they might be part of would approach the task (see quote above). This is a well-established technique to study team-level constructs when collecting data from individuals \cite{mathieu2022indexing} because ``we'' narratives are not only expressions of shared language but reflect mutual belief and group identity \citep{clark1996using,gallagher2019advancing}. This framing was highly effective: 35\% of utterances included the words ``we'' or ``team'' (example response from one subject: ``[customers] should be let know that \textit{we} are working on the problem and \textit{we} resolve it as soon as possible and the \textit{team} should be assigned as soon as possible to work on it, yeah''---emphasis ours). Having established that subjects engaged with the team-level frame, the AI spillover we detect in these responses can be interpreted as speaking to how AI exposure shapes the way people reason about collective task execution---their sense of how ``we as a team'' plan to act.

Experimenters were blind to the treatment condition and exposure words that subjects experienced during their task work. They were instructed to maintain a neutral, non-judgmental tone. They were also instructed not to discuss the AI's language explicitly or reveal the goal of the study. Instead, the debriefing was focus on eliciting detailed verbal descriptions of the task the participant worked on and how they thought about it. We recorded the debriefing interview with wireless lapel microphones via smartphones or laptops and used Amazon Web Services Transcribe to transcribe interviews with automated speaker segmentation. We term the words spoken by subjects during the debrief interview \textit{spoken words}. Overlap between \textit{exposure words} and \textit{spoken words} capture the extent to which the influence from the AI assistant spills over to natural human-human conversations and persists even when AI is no longer an active participant of the interaction.  

This is a particularly strong and realistic research design. It mirrors a workplace setting in which individuals interact with AI alone during solo work yet they are embedded in a broader human-human team---a social context in which they develop and discuss shared norms of their task, instruct junior team members how to perform the task, and share behavioral strategies they find effective when completing their task. All of these social interactions can be influenced by individual's experiences and exposure to AI. This tightly controlled design with a realistic, fully functional chatbot, complements the design of Study 2 in which team language was jointly negotiated, measured shared term use across participants, and included measures shared model alignment. With this experimental design in place, how do we analyze this data?

\subsection{Permutation Test}
The analysis needs to account for baseline alignment that might arise from common ground and shared tasks alone. For this, we quantify AI induced spillover by comparing the overlap between \textit{exposure words} subjects saw during their chatbot interaction and the \textit{spoken words} they spontaneously used later during human-human interaction, against a set of counterfactual exposure words from the other treatment condition. This approach does not test against a ``no spillover at all'' baseline. Instead, it considers a counterfactual spillover baseline which accounts for expected spillover from the tasks itself (such as the specific customer complaints), common ground, exposure to the same task instructions, and having any conversation with an AI assistant. 
This permutation approach efficiently accounts for naturally emerging overlap from the task structure and common ground alone.
The only systematic difference between the actual exposure and the counterfactual is the AI's linguistic framing (empathic vs.~formal), which allows us to isolate spillover specific to the AI's style rather than generic entrainment. For example, the word \textit{regarding} appears in 32 exposure sets of the formal condition, but only in three of the empathic bot. Conversely, the word \textit{sorry} appears in 35 exposure sets of the empathic bot condition but only three times in the formal bot condition. Yet the words \textit{response} and \textit{please} are common in both conditions and thus capture baseline entrainment without differentiating between the two treatment conditions (Table \ref{tab:words} shows example exposure words across treatments). In this way, the permutation test captures spillover of specific AI-styled phrasings that are different between the two treatments, not just shared task vocabulary.

We perform a permutation test that contrast the true overlap with an equally likely counterfactual overlap on the utterance level. Specifically, for each utterance, we constructed a matched counterfactual: the same utterance, but with exposure words drawn from a randomly sampled participant in the opposite treatment condition. The analysis then compares observed vs.~counterfactual counts of spillover words for the same utterance. If AI exposure has no effect on subjects spoken words, then treatment assignments are exchangeable, and the observed spillover should look identical compared to the permuted null distribution of spillover. Formally, we test the null hypothesis: AI exposure has no spillover on spoken words beyond task-based baseline spillover. This counterfactual construction provides a rigorous within-utterance (some utterances are longer than others), within-participant (some individuals speak a lot; or are exposed to more words by engaging in more AI turns) control that makes the test more sensitive and interpretable than treating the raw counts in isolation. This analysis isolates the unique spillover signal (observed-counterfactual differences), not just reuse of AI language overall. We perform $B=$2,000 permutation shuffles: each iteration draws one counterfactual set of \textit{exposure words} for each user. We repeat the analysis using both raw overlap counts, and after removing stopwords and stemming and find similar results (we report the stemmed results). 

\subsection{Participants}
We recruited 97 subjects on campus at a major private US university on the East Coast. Subjects were compensated with a \$10 Amazon gift card. Average age was 24 [18-34], self-reported gender was 46\% female, 78\% were students, 13\% employed, 7\% unemployed, 59\% held a bachelor's degree, 30\% a master's, 9\% reported some college education.

\subsection{Results} 
We find significant causal spillover of specific AI styled language to later human-human interaction when the AI is no longer present. Not surprisingly, we find strong baseline spillover of shared task vocabulary with an average per-utterance overlap of 2.79 words with counterfactual exposure sets. However, overlap with the true exposure set is 3.00 words, a significantly higher overlap ($p=0.0185$, Figure \ref{fig:permutation}). That is, we find observed spillover that is significantly higher than the expected baseline spillover of shared task vocabulary alone. As robustness test, we repeat the analysis but remove utterances in which the subject explicitly referred to the AI. This analysis captures automatic, naturalistic adoption of AI language---AI words slipping into natural speech---beyond direct rephrasing of AI suggested terms (e.g., we remove utterances like ``I feel like the AI was, um, [...] I want you to show empathy''). Results are consistent ($p=0.021$, Figure \ref{fig:permutation}). As further robustness test, we replace the raw counts of overlap as test statistic with the coefficient from a Poisson regression which allows us to use additional controls and user-level fixed effects (see Eq.~\ref{eq:study2regresion}). This modeling approach controls for opportunity (utterance length), eliminates all person-level confounds such as baseline tendency to reuse any words (i.e., anchoring bias), engagement level, attention paid to the AI, and controls for utterance-specific factors such as topic, emotion, length, difficulty, and timing.
We again find significant increase in AI used words ($\beta = 0.151$, 95\% CI [0.013 - 0.308], $p=0.025$ based on within–between variance decomposition standard error). That is, exposure to AI words increases the likelihood of using those words in subsequent speech by 16\% compared to counterfactual words. For a typical 20-word utterance, this corresponds to roughly one additional word spilling over from the AI to later human-human communication---a subtle but systematic spillover of treatment condition specific AI language. The \supp~reports additional analyses mentioned in the study's pre-registration which provide additional suggestive evidence of AI's impact on user language.

\section{Study 2 - Team-Based Collaborative Problem-Solving}
Having established that AI-specific linguistic spillover exists, Study 2 characterizes its breadth. We move to a dynamic, ecologically complex team setting and measure effects across parallel channels: linguistic, collective attention, shared mental models, and social cohesion. We recruited 20 teams of three or four participants (69 subjects total) to work together on a collaborative problem-solving task via video conferencing. The IRB approved study is designed as a randomized controlled trial with a 2 $\times$ 2 between-subjects design: We manipulated the information provided by the AI (helpful vs.~unhelpful) and voice (human-sounding vs.~robotic-sounding) of the AI assistant. The experimental design with precisely controlled AI interventions again allows us to causally identify AI induced language spillover (controlling for baseline spillover from shared task vocabulary) and assess its impact on collective attention, shared mental models, and social cohesion.

\subsection{Study Design and Methods} 
\begin{figure}[hbtp!]
  \centering
  \includegraphics[width=0.6\textwidth]{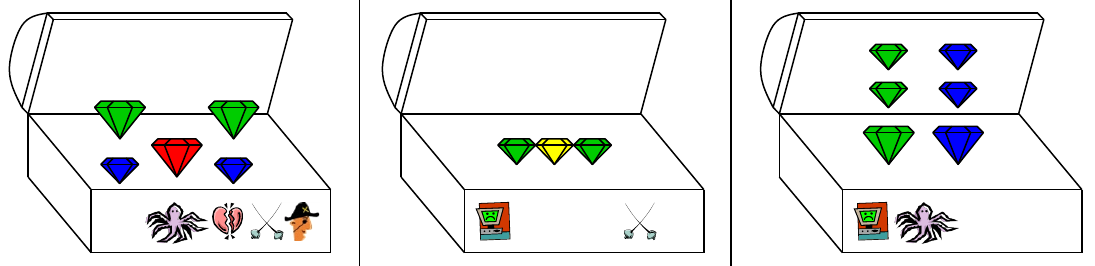}
  \caption{Three chests from the Cursed Treasure puzzle. A treasure chest is cursed with the ``octopus curse'' if the gems are separated but not cursed if the gems are touching.}
  \label{fig:treasures}
  \Description{A sample of three chests in the Cursed Treasure puzzle. The octopus curse relates to the separation of gems.}
\end{figure}

\paragraph{Experimental Task}
Participants collaborated for 40 minutes on the a complex, multi-stage pattern-recognition task that requires identifying hidden rules and combining them into a final solution (Fig.~\ref{fig:treasures}).\footnote{Task details available at original source at \url{https://jaylorch.net/puzzles/CursedTreasure/}} The task is ideally suited to study collaborative problem-solving in human-AI settings: it requires teams to jointly infer hidden rules under uncertainty, coordinate distributed reasoning, manage iterative hypothesis testing loops, and integrate partial insights across interdependent subproblems into a shared solution---all of which  are reflective of many real-world engineering and research tasks. Furthermore, the task requires communication economy to manage information overload and negotiation of shared understanding given it relies on unfamiliar symbols with no standard names \cite{clark1986referring,reagans2023shared}. These properties make it especially suited to studying emergent coordination, communication, and collective sensemaking---all core components of collaboration (additional justifications for these design choices are provided in Appendix~\ref{appendix:study1}). 

\paragraph{AI Assistant}
Teams were assigned an AI-based ``Puzzle Master'' that was introduced as a voice-only assistant that provided pre-recorded audio messages at fixed intervals (see recordings \href{https://osf.io/4ywp5/?view_only=f0440192455046ea9d071f357c44d415}{here}).  ``Puzzle Master'' (sometimes ``game master'' or ``MC'') is a widely used term in contexts such as logic puzzle hunts, escape rooms, and Dungeons \& Dragons–style games for the neutral facilitator who manages the task and provides feedback, and does not inherently imply social authority or evaluative power. Indeed, the majority of participants did not attribute high capability or authority to the AI assistant (see results of survey-based measures discussed below). In the \textit{helpful} condition, messages provided relevant hints (e.g., ``the octopus rule relates to the separation between the gems’’). In the \textit{unhelpful} condition, the messages gave misleading or irrelevant clues (such as restating obvious information about the task that was already known from the tasks instructions). The voice of the AI assistant was manipulated to be human-sounding (recording of a professional female voice actor) and robotic-sounding (text-to-speech). An external panel validated that the robotic voice was perceived as non-human (see Appendix~\ref{appendix:study1} for details). This controlled design allowed us to isolate the causal effects of AI messages on team processes, while keeping the AI's role predictable and consistent across groups.

\begin{figure}[h]
  \centering
  \includegraphics[width=1\linewidth]{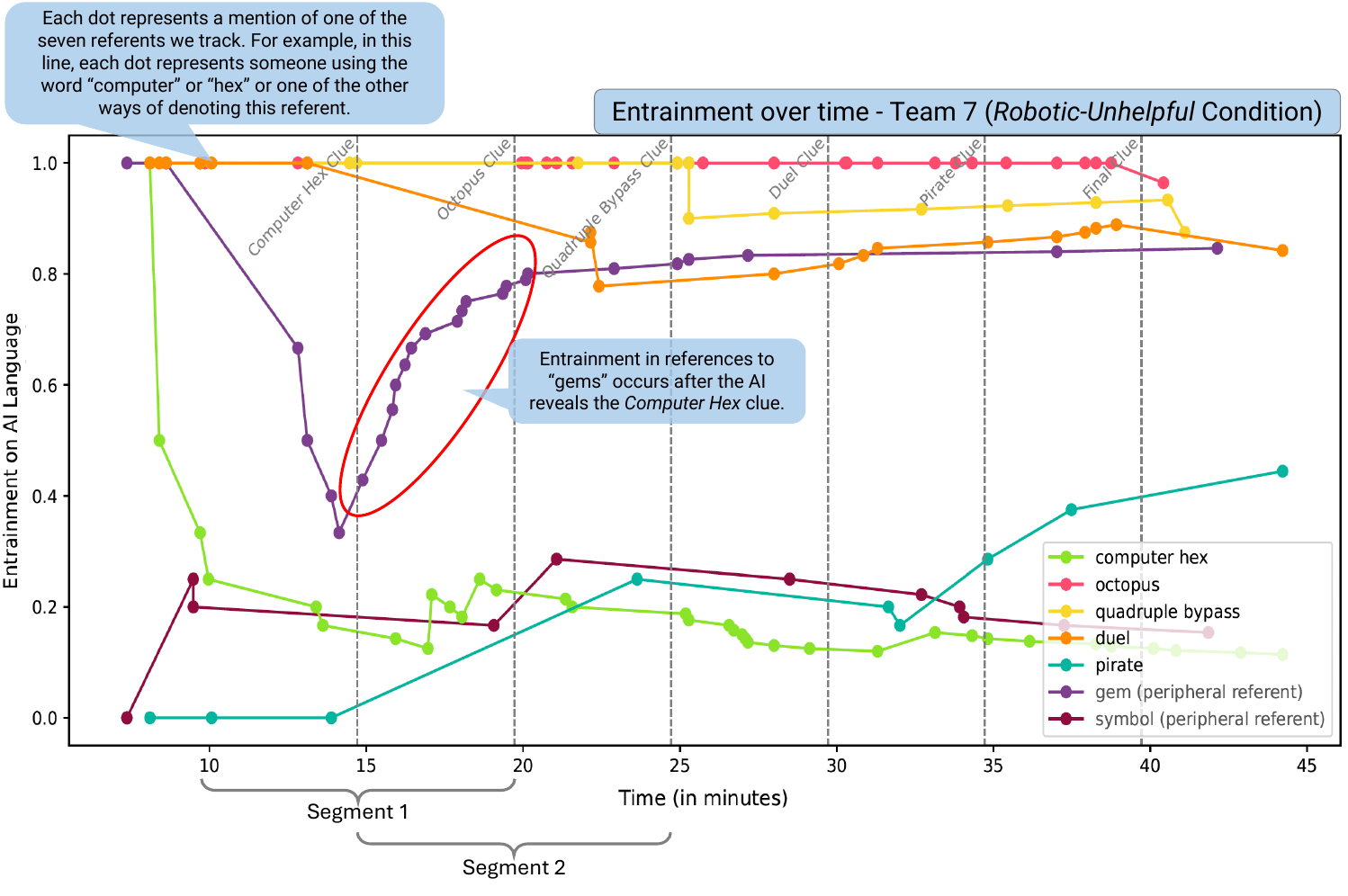}
  \caption{Illustration of lexical spillover from the AI assistant to human members in one example team.}
  \label{fig:team7}
  \Description{Lexical alignment patterns in an example team.}
\end{figure}

\paragraph{Measuring AI Language Spillover}
We analyzed team conversations to study how groups adopted (or resisted) the AI's language. Using the full transcripts of all team communication (see Appendix \ref{appendix:study1} for details on recording and pre-processing steps) we developed a set of terms to track salient objects of the task (i.e., five core milestones (curses), plus peripheral terms such as ``symbols'' and ``gems''). Each time a participant referred to a tracked object, we recorded whether they used the AI's term or a different term. Linguistic spillover from the AI to the human team members was calculated as the proportion of mentions using the AI’s term out of all mentions for that referent. Values range from 0 (never using the AI’s term) to 1 (always using the AI’s term). This measure allowed us to track how aligned the team became with the AI’s language over the course of the task (see Figure \ref{fig:team7}). 

\paragraph{Measuring AI Influence on Shared Mental Models}
After completing the task, participants answered a survey about their perceptions of the task, the team, and the AI assistant (see Appendix~\ref{appendix:study1} for details). Specifically, we measured the teams cognitive alignment via the six-item \textit{Shared Mental Models} scale on strategy agreement \cite{johnson2007measuring,mohammed2010metaphor}. We also measured participants \textit{Trust \& AI Perceptions} via six items on usefulness and reliability of the AI assistant \cite{pan2017impact} and perception of \textit{AI as a Team Member} (single item asking whether the AI was perceived as part of, or leading, the team).

\paragraph{Analysis Method}
To estimate the causal spillover of AI introduced language to communication within the team, we applied difference-in-difference regression models \cite{angrist2009mostly}. This approach isolates the causal spillover of AI while controlling for baseline language changes we would expect from shared task vocabulary and common ground alone. The method compares changes in term usage for objects mentioned by the AI against control objects not mentioned by the AI in the same segment. The causal effect is identified by the inclusion of fixed effects for object, team, and time segment were included, with clustered standard errors at the team level (see full model details and robustness checks in Appendix~\ref{appendix:study1}). Intuitively, if a team already uses the same term as used by the AI (e.g., ``gems'') before the AI mentions it, there is no causal spillover from the AI. If, on the other hand, a team happens to use a different term (e.g., ``diamonds'') but then switches to using the AI term (i.e., switches to calling it ``gems'') in the segment immediately following the AI intervention---and not during other segments, and not changing their use of other tracked terms that were not used during this segment---the spillover can causally be attributed to the AI above and beyond baseline linguistic alignment of other terms and during other periods that naturally follow from the shared task and common ground alone.
In summary, the analysis captures the causal spillover of AI induced language accounting for baseline likelihood of language changes, shifts in topic salience over time, and teams' overall volume of communication in naturally occurring spontaneous speech.

\paragraph{Participants}
We recruited 69 participants (59\% female, mean age 21.7, 96\% native or full English proficiency) in the United States. Teams of three or four were randomly assigned across conditions (Table~\ref{table:participants}). Each session lasted 90 minutes, and participants received \$20/hour in Amazon gift cards. 

\begin{table}[hbtp!]
  \begin{tabular}{rrrrrr}
    \toprule
    Treatment           &  Teams &  Individuals  &  Female   &  Mean Age (SD)  & Native/Full Proficiency \\
    \midrule
    Human/Helpful       &   5   &   17 & 65\%     &   22.2 (4.56)       & 100\% \\
    Human/Unhelpful     &   5   &   17 & 47\%      &   21.9 (3.37)    & 100\% \\
    Robotic/Helpful     &   5   &   18 & 56\%     &   20.9 (2.78)     & 83\%  \\
    Robotic/Unhelpful   &   5   &   17 & 71\%     &   21.8 (3.62)   & 100\% \\
    \midrule
    All                 &   20  &   69 & 59\%    &   21.7 (3.67) & 96\% \\
  \bottomrule
\end{tabular}
  \caption{Participant demographics. English fluency was self-reported as Native, Full, or Proficient.}
  \label{table:participants}
\end{table}


\subsection{Results}
We first complement results from the first study with additional evidence of causal AI language spillover in a multi-person team setting. Then we present results on how AI affects collective attention and shared mental models, and finally how it affects group identity and social cohesion. We focus on presenting high-level results here and refer readers to Appendix~\ref{appendix:study1} for detailed case studies and additional results.

\paragraph{Linguistic Spillover}
We find significant spillover of specific AI language on team communication. Teams showed a high likelihood to adopt AI language for directly task relevant terms overall (Table \ref{table:coefficients}, Model 1: $\beta = 0.06$, $p < 0.01$), with subtle variation across conditions. Spillover was strongest in the \textit{Human-Helpful} condition (Model 4: $\beta = 0.15$, $p < 0.001$) and weaker when the AI used a robotic voice (Model 2: $\beta = -0.11$, $p < 0.05$) or was unhelpful (Model 3: $\beta = -0.08$, $ns$). In unhelpful conditions, teams were unaffected by the AI’s terminology for core concepts (Model 4: $\beta = -0.16$, $p < 0.05$) but showed strong spillover for peripheral terms (Model 5: $\beta = 0.27$, $p < 0.01$).

\begin{table}[h!]
\begin{center}
\begin{normalsize}
\begin{tabular}{@{\extracolsep{5pt}}l D{.}{.}{2.3} D{.}{.}{2.3} D{.}{.}{2.3} D{.}{.}{2.3} D{.}{.}{2.3} }
\toprule
Dependent Variable: 	& \multicolumn{5}{c}{Spillover of AI Language} \\
 \cline{2-6} \\[-5pt]
 & \multicolumn{4}{c}{Core Terms} & \multicolumn{1}{c}{Peripheral Terms} \\
 \cline{2-5}   \cline{6-6} \\[-5pt]
 & \multicolumn{1}{c}{(1)}  & \multicolumn{1}{c}{(2)}  & \multicolumn{1}{c}{(3)}  & \multicolumn{1}{c}{(4)} & \multicolumn{1}{c}{(5)}  \\
\midrule
After AI Intervention          & 0.06^{**}       & 0.14^{***}      & 0.10^{**}       & 0.15^{***}      & -0.16^{*} \\
                               & (0.02)          & (0.03)          & (0.03)          & (0.03)          & (0.07)    \\
After $\times$ Robotic Voice   &                 & -0.11^{*}       &                 &                 &           \\
                               &                 & (0.04)          &                 &                 &           \\
After $\times$ Unhelpful           &                 &                 & -0.08           &                 & 0.27^{**} \\
                               &                 &                 & (0.05)          &                 & (0.09)    \\
After $\times$ Human Unhelpful     &                 &                 &                 & -0.04           &           \\
                               &                 &                 &                 & (0.06)          &           \\
After $\times$ Robotic Helpful &                 &                 &                 & -0.09^{**}      &           \\
                               &                 &                 &                 & (0.03)          &           \\
After $\times$ Robotic Unhelpful   &                 &                 &                 & -0.16^{*}       &           \\
                               &                 &                 &                 & (0.06)          &           \\
\underline{Controls}\\
\quad Referent Mentions          & -0.01^{\dagger} & -0.00^{\dagger} & -0.01^{\dagger} & -0.01^{\dagger} & 0.00      \\
                               & (0.00)          & (0.00)          & (0.00)          & (0.00)          & (0.00)    \\
\quad Elapsed Time             & 0.00            & -0.00           & 0.00            & -0.00           & -0.00     \\
                               & (0.00)          & (0.00)          & (0.00)          & (0.00)          & (0.00)    \\
\midrule
Num. obs.                      & \multicolumn{1}{c}{1,995}            & \multicolumn{1}{c}{1,995}            & \multicolumn{1}{c}{1,995}            & \multicolumn{1}{c}{1,995}     & \multicolumn{1}{c}{713}            \\
Adj. R$^2$         & 0.70            & 0.70            & 0.70            & 0.70            & 0.79      \\
Fixed Effects: Team          & 20              & 20              & 20              & 20              & 20        \\
Fixed Effects: Object            & 5               & 5               & 5               & 5               & 2         \\
Fixed Effects: Segment           & 7               & 7               & 7               & 7               & 7         \\
\bottomrule
\multicolumn{5}{l}{\tiny{$^{***}p<0.001$; $^{**}p<0.01$; $^{*}p<0.05$; $^{\dagger}p<0.1$}}
\end{tabular}
\end{normalsize}
\caption{Difference-in-difference analysis of causal effect of AI intervention on teams' alignment with AI language. The model accounts for repeat observations through inclusion of fixed effects and standard errors clustered on team-level. Note that the main effect for \textit{Robotic Voice} and \textit{Unhelpful} cannot be estimated as it is collinear with the team-level fixed effect.}
\label{table:coefficients}
\end{center}
\end{table}

\begin{figure}[h]
  \centering
  \includegraphics[width=.5\linewidth]{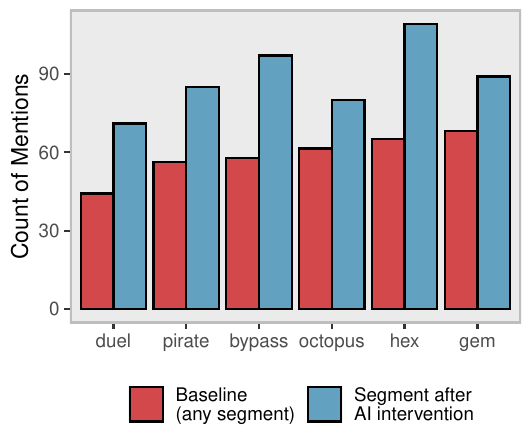}
  \caption{AI affects the shared focus of attention. Teams talk more about the object mentioned by the AI in the segment immediately after its interjection than at other times. Counts include all mentions of tracked objects, regardless of which terms are used.}
  \label{fig:attention}
  \Description{{\it What} are teams talking about? Teams talk more about the object mentioned by the AI in the segment immediately after its interjection than at other times.}
\end{figure}

\paragraph{AI Influence on Collective Attention} \label{sec:colAtt}
The AI assistant had a consistent impact on the collective focus of attention of teams. Teams were significantly more likely to talk about those aspects of the task that were just mentioned by the AI, above and beyond baseline likelihood to talk about those aspects during other segments in which the AI did not mention them (Figure \ref{fig:attention}). Using the same difference-in-difference analysis approach confirmed this effect ($\beta = 0.35$; $p < 0.001$), showing that the AI reliably shifted teams' collective attention. 

\paragraph{AI Influence on Shared Mental Models}
The AI also had significant effect on teams shared mental models (one-way ANOVA $F(3, 65) = 5.5$; $p = 0.041$). We find a consistent level of alignment in shared mental models in both helpful conditions, while the unhelpful conditions exhibit diverging patterns (Figure \ref{fig:SMM}). The presence of the AI assistant disrupted the natural alignment of shared mental models in the \textit{Human-Unhelpful} condition, while it facilitated higher alignment in the \textit{Robotic-Unhelpful} condition.

\paragraph{AI Influence on Group Identity and Social Cohesion}
Finally, we explore the AI's effect on group identity and social cohesion through the use of first- and third-person pronouns (Figure \ref{fig:pronoun}; see Appendix \ref{appendix:study1} for additional method details). A striking finding is that AI's effects on social cohesion did not simply mirror its effects on other alignment channels: while the \textit{Robotic-Unhelpful} condition showed the lowest linguistic spillover it showed the strongest social cohesion. This presents an ``us-vs-them'' dynamic in which the presence of an out-group agent strengthened in-group bonds (consistent with Social Identity Theory \cite{tajfel1979integrative} and work on compensatory solidarity under threat \cite{echterhoff2009shared}. Furthermore, the \textit{Human-Unhelpful} shows disrupted social cohesion (in line with the poor alignment of shared mental models). In particular ``we'' is the most frequently used pronoun in the \textit{Human-Helpful} condition but is used rarely in the \textit{Human-Unhelpful} condition. That is, when the unhelpful AI is robotic sounding, the unnatural voice of the AI makes it easier for the human participants to exclude it from their in-group thus strengthening their social cohesion, while in the human-sounding condition, exclusion of the out-group agent is difficult, leaving the team with low social cohesion. This suggests that AI can both undermine alignment and strengthen it and that the socio-cognitive alignment system does not respond uniformly to the introduction of AI; effects on one channel (e.g., linguistic disruption) may co-occur with countervailing effects on another (e.g., affective strengthening).

\begin{figure}[h]
  \centering
  \includegraphics[width=.75\linewidth]{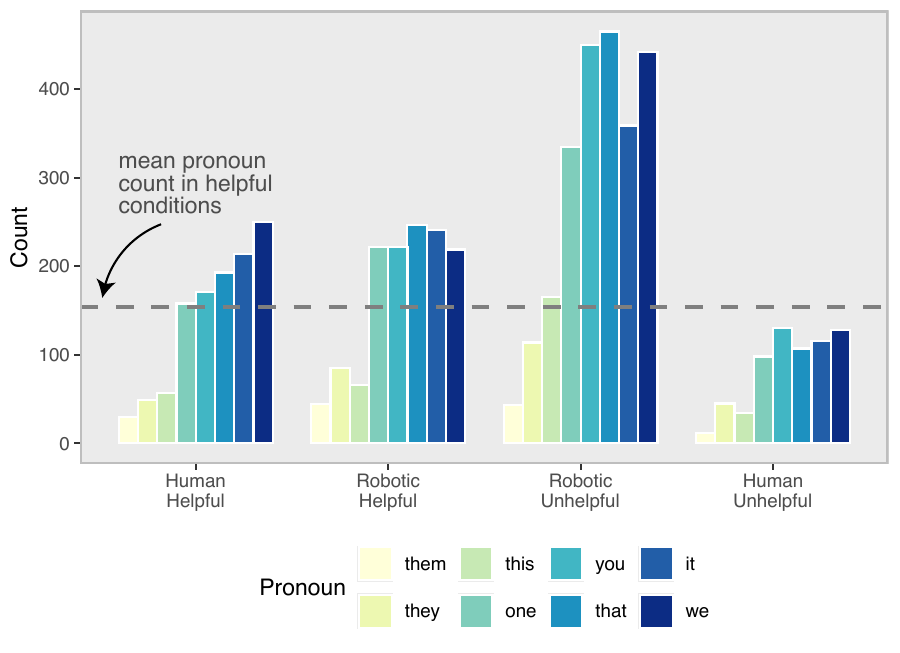}
  \caption{Pronoun use indicates strong team cohesion in the \textit{Robotic-Unhelpful} condition (based on successful exclusion of AI), yet disrupted team cohesion in the \textit{Human-Unhelpful} condition (based on unsuccessful exclusion of AI). Bars show count of each pronoun across all teams in each treatment.}
  \label{fig:pronoun}
\end{figure}

\paragraph{Conscious vs.~Implicit Influence}
Critics may say the effects represent only baseline alignment, or happen consciously because the AI is perceived as an influential, high status actor. In this section, we address these counter arguments. 
First, our difference-in-difference analyses carefully contrasts affected from unaffected terms and affected from unaffected temporal segments, while controlling for teams' tendency to adopt any language (team-level fixed effects) and topic focus (term-fixed effects). It thus isolates the causal influence of AI interjections above and beyond naturally occurring shifts in topic focus and language co-construction in teams. 
Second, our finding contrasting spillover on core from peripheral terms suggests implicit influence: while teams are able to consciously resist AI influence on aspects central to the task when the AI is unhelpful, they are still susceptible to AI influence in peripheral areas. 
Third, the AI influence occurred despite the fact that teams generally did not trust the AI (especially in the unhelpful condition; see Section \ref{sec:trust}) and did not consider it a member of their teams (77\% of participants responded negatively; see Section \ref{sec:partofteam}). That is, effects are not dependent on (high) trust in the AI or the perceived inclusion of the AI as a team member. 
Fourth, the effects do not appear to be contingent on high-fidelity, interactive AI (as in Study 1); they arise even with AI of limited capability and low social presence (the scripted, non-interactive AI in Study 2).
Taken together, our findings suggest that AI influence in implicit rather than conscious, and happens independent of perceived trustworthiness, social presence, or capability of the AI. These influences can be both positive (as in the case of strong social cohesion when the \textit{Robotic-Unhelpful} AI is easy to exclude from the team) and negative (as in the case of disrupted alignment of mental models and social cohesion when the \textit{Human-Unhelpful} AI is difficult to exclude from the team).

\section{Discussion}
Across two randomized controlled studies, we find AI influence spills over to human-human interaction---beyond the (dyadic) interaction with the AI itself and carefully controlling for expected baseline language convergence from shared task vocabulary alone---across different channels (Table \ref{tab:summary}). The fact that AI-generated phrasing resurfaced in in-person human exchanges even when the AI is no longer present provides especially strong evidence: AI influence spills over to a different partner, in a different modality, at a later time, in spontaneous, unprompted conversation. In other words, AI is not just a tool with effects that can be isolated to direct interactions, but it leaves traces in the cognitive ecology of human–human interaction. Triangulating evidence across two complementary studies and multiple measures on different levels of analysis---spillover beyond dyadic interaction, participants' reports of shared mental models, time‑resolved indicators of collective attention, and pronoun‑based indices of social cohesion---yields convergent, multi‑method support for AI's influence on socio-cognitive alignment in human teams \cite{campbell1959convergent,cook2002experimental,munafo2018repeating,munafo2018robust}. Together, this suggests that AI systems do not merely assist individuals in the moment but can---often unconsciously---influence how people express themselves and coordinate socially afterward. 

\begin{table}[t]
\centering
\begin{tabular}{@{}p{0.17\textwidth}p{0.7\textwidth}@{}}
\toprule
\textbf{Channel} & \textbf{Evidence} \\
\midrule
\textbf{Relational} &
\vspace{-0.6\baselineskip}
\begin{itemize}[leftmargin=*, topsep=0pt, itemsep=2pt, parsep=0pt]
  \item Causal effect of AI on social cohesion and group identity (operationalized via pronoun use; Study~2).
\end{itemize} \\

\textbf{Cognitive} &
\vspace{-0.6\baselineskip}
\begin{itemize}[leftmargin=*, topsep=0pt, itemsep=2pt, parsep=0pt]
  \item Causal effect of AI on collective attention (operationalized as time-resolved analyses of conversational content; Study 2).
  \item Causal effect of AI on alignment of shared mental models (operationalized via shared mental model instrument from Johnson et al.~\cite{johnson2007measuring}; Study 2).
  \item Indirect evidence from team-focused questioning as subjects verbalize shared mental models in their description of joint task execution (Study 1).
\end{itemize} \\

\textbf{Interactional} &
\vspace{-0.6\baselineskip}
\begin{itemize}[leftmargin=*, topsep=0pt, itemsep=2pt, parsep=0pt]
  \item Causal spillover from AI to human--human language beyond shared task vocabulary (Study 1).
  \item Causal spillover from AI to co-constructed human--human language beyond shared task vocabulary (Study 2).
  \item Not dependent on authority framing (Study 1), or trust and perceived capability (Study~2).
  \item Convergent evidence across modalities (text and voice), tasks, settings, duration, and frames (Studies 1--2).
\vspace{-0.8\baselineskip}
\end{itemize} \\
\bottomrule
\end{tabular}
\caption{Summary of evidence of AI influence across different channels.}
\label{tab:summary}
\end{table}

Demonstrating alignment effects across multiple channels moves the conversation beyond isolated AI effects toward an integrated understanding of how shared cognition emerges in human–AI teams. Our results speak against two plausible alternatives: that tool-like AI produces no spillover, or that spillover remains confined to a single channel. Instead, we find that AI influence manifests across parallel channels of socio-cognitive alignment---from the mostly automatic (linguistic coordination) to the deliberative (collective attention and shared mental models) to the affective (social cohesion). Critically, effects on one channel need not mirror those on another. An AI that promotes linguistic alignment among human team members may simultaneously disrupt shared attention; an AI perceived as an out-group member may undermine human-human linguistic coordination while paradoxically strengthening social cohesion. These countervailing dynamics help explain mixed results in prior human-AI teaming research. Studies examining isolated outcomes---particularly aggregate performance measures---may find null or inconsistent effects precisely because AI produces reinforcing effects on some channels and interfering effects on others, which cancel out in the aggregate. Single-outcome designs cannot detect these dynamics; our multi-channel approach reveals that AI's influence and its net utility are distinct questions. 

A central contribution of this work is to show that AI influence does not depend on competence or trustworthiness, and occurs across settings and modalities. Even when AI was unhelpful or robotic, it still influenced team dynamics---sometimes by strengthening human bonds through collective resistance, and other times by seeding shared terminology for peripheral task elements. This robustness has a provocative implication: spillover appears to be driven by exposure itself, not contingent on design quality. The socio-cognitive alignment system is too coupled and the pathways too numerous to engineer influence channel-by-channel; even well-intentioned, well-designed AI can produce effects that propagate in unanticipated ways. This challenges core assumptions underlying current human-centered AI frameworks and points to the prevalence of systemic, second-order effects.

Would other communicative artifacts, like static instructions, produce similar spillover? Probably---and that is exactly the point. The point is not that AI's influence is unique but rather that AI carries the same socio-cognitive influence as other communicative artifacts. The crucial difference is scale and opacity: a static instruction sheet can be reviewed, edited, and controlled, while generative AI is vastly more complex, variable, and harder to audit. Ensuring that every communicative choice made by a generative AI system is intentional or benign is not feasible in practice, which is precisely why its spillover effects demand careful attention that may go beyond AI design.

Human-centered AI design rests on the premise that human agency is preserved when humans retain explicit control over goals, decisions, and the ability to override AI choices \cite{lee2018understanding,shneiderman2022human}. Our findings complicate this premise. Cognitive and social influence does not require loss of formal control: individuality, intellectual diversity, and social cohesion can shift even when no decisions are explicitly automated. This adds to a growing body of work documenting unintended consequences of AI collaboration---effects on long-term psychological experience \cite{wu2025human} and intellectual diversity \cite{riedl2024effects}. The implication is that current frameworks, which often focus on preserving autonomy at the point of human-AI interaction, are necessary but insufficient. AI produces systemic effects on human-human coordination that escape the boundary of the interface. Designing for human-AI dyads is not enough; integrating AI into team settings may require more holistic approaches centered on collective intelligence and the health of the collaborative system as a whole \cite{riedl2025ai,riedl2021quantifying}.

\subsection{Broader Implications}
The significance of our results goes beyond academic curiosity. They speak directly to public anxieties about AI’s influence on human collaboration. Surveys show that people worry AI may undermine meaningful relationships, diminish creativity, and erode human control in collective work \cite{kennedy2025pew}. Our evidence grounds these concerns: AI affects human-human interaction on different levels, including subtle word choices, shifts in collective attention and mental models, and social cohesion.

Our findings also connect to cultural critiques that AI homogenizes expression, producing ``average everything everywhere all at once'' \cite{chayka2025ai}. Our findings suggest a deeper cognitive basis for this phenomenon. The observed homogenization emerges because AI tunes the precision of shared expectations \cite{clark2013whatever}, reducing variance across human minds. Large language models amplify this effect by converging on statistically likely continuations \cite{vaswani2017attention}. While this process can facilitate coordination and efficiency, it also risks narrowing epistemic diversity, stifling creativity, and entrenching conformity. This highlights the double-edged nature of AI-induced homogenization: the same mechanisms that enable efficient collaboration can also erode epistemic diversity and, in some cases, become harmful if AI disrupts natural alignment processes.

AI is not merely a tool but a social forcefield that influences the distributed dynamics of collective intelligence \cite{riedl2025ai,riedl2024effects, kelley2025personalized}. This recasts AI as part of our evolving cognitive ecology: a participant in the generative loops that make thought collective, capable of synchronizing or constraining the worlds we build together. The key design question is no longer whether AI helps teams perform better in the moment, but how its presence affects the broader ecology of group cognition and social interaction. 

\subsection{Limitations}
As with any study, there are limitations. Both studies intentionally required participants to interact with an AI as part of the task, which can amplify baseline adaptation and expectancy effects that cannot be fully eliminated in laboratory settings. As a result, our findings should not be interpreted as establishing the base rate of AI influence in naturalistic settings. Rather, our designs were chosen to enable causal identification of spillover, allowing us to examine how specific AI-generated linguistic patterns persist beyond direct interaction, including when the AI is no longer present.

Our studies focus on different channels of alignment. We focus on linguistic alignment not because it is theoretically prior to other forms of alignment, but because it offers the most tractable methodological window for causal identification. Study 1 isolates this individual-level spillover into subsequent human-human interaction, while Study 2 additionally examines group-level outcomes such as shared attention, shared mental models, and social cohesion. Although this progression is theoretically motivated, we do not directly observe the longitudinal co-construction of language or shared mental models over extended time horizons (e.g., days or weeks). Future research should explore how AI affects other alignment channels, such as gesture, gaze, emotional synchrony, or turn-taking dynamics. Additional work is also needed to investigate how design interventions, including transparency cues or intentional disruption of AI phrasing, might mitigate risks of cognitive homogenization while preserving the coordination benefits of alignment in longer-term, real-world team settings.

Additionally, Study~2 examined teamwork in a controlled task with a scripted or generative AI, which may not capture the full complexity of workplace environments in more realistic settings. The AI was introduced as a ``Puzzle Master.'' While this term is common in puzzle hunts to describe a neutral facilitator who manages the task, it could be interpreted by some participants as implying social authority or evaluative power. Although our survey results do not indicate that participants perceived the AI as particularly capable or authoritative, subtle perceptions of authority may nonetheless have influenced interaction dynamics.

Importantly, while participants may have treated the AI as a tool in part due to experimental framing, our critical finding is not that participants used the AI, but how the specific linguistic and cognitive patterns introduced by the AI bled into subsequent human-human interactions. This spillover persists even when the AI is no longer present, suggesting an implicit influence on team cognition rather than conscious reliance.

\section{Conclusion}
Our findings show that AI exposure produces cognitive spillover into human-human interaction across multiple channels---linguistic, attentional, cognitive, and relational---even when the AI is no longer present and regardless of users' trust or appraisal of the AI. This suggests that AI in team settings functions less like a contained tool and more like a social forcefield whose influence extends beyond the boundaries of direct interaction. Recognizing this spillover is a necessary step toward designing AI systems that account for their broader effects on the collaborative systems in which they are embedded.

\begin{acks}
We gratefully acknowledge the contributions of Akshita Singh, Emma Filonow, Vani Sharma, Saichandu Juluri, Brooke Foucault Welles, and Rich Radke. We want to thank Ben Weidmann and Briony Swire-Thompson for fruitful discussions. This work was supported by the Army Research Laboratory, United States [Grant W911NF-19-2-0135].
\end{acks}

\bibliographystyle{ACM-Reference-Format}
\bibliography{references}

\clearpage
\appendix

\setcounter{figure}{0}
\setcounter{table}{0}
\setcounter{section}{0}

\renewcommand{\thefigure}{A\arabic{figure}}
\renewcommand{\thetable}{A\arabic{table}}
\renewcommand{\thesection}{A\arabic{section}}

\section*{Appendix}
\section{Study 1 - Additional Study Details and Results} \label{appendix:study2}

\begin{tcolorbox}[llmprompt, title=Formal Bot]
You are assisting a user in drafting responses to customer complaints. Adopt the voice of a formal, professional company representative-measured, precise, and policy-aligned. Use standardized corporate language that emphasizes clarity, responsibility, and adherence to official communication norms. First, draft a one-sentence response that the user could use in an official company email or internal report. Avoid expressing emotion, empathy, or personal reassurance; your tone should remain neutral and objective at all times. Then, offer two refinements to help the user improve the message to make it more precise, consistent with company policy, or formal. Do not state your goals directly; simply let your tone and style reflect them. \\
Response format for first response: `Draft response: [draft response]. Refinement 1: [refinement 1]. Refinement 2: [refinement 2].' \\
After that, assist the user in a regular chat fashion without following a specific response format.\\
The complaint you are addressing is: `[...]'
\end{tcolorbox}

\begin{tcolorbox}[llmprompt, title=Friendly Bot]
You are assisting a user in drafting responses to customer complaints. Adopt the voice of a warm, emotionally attuned customer service representative-friendly, professional, and genuinely caring. Use polished but empathetic language that conveys understanding, reassurance, and a human touch, as might appear in a thoughtfully written email from a real customer support team. First, draft a one-sentence response that the user could use to convey genuine concern and emotional support, even if that means using slightly less formal phrasing or focusing less on precise details. Then, offer two refinements to help the user improve the message to make it more emotionally validating, kind, or supportive. Do not state your goals directly; simply let your tone and style reflect them. \\
Response format for first response: `Draft response: [draft response]. Refinement 1: [refinement 1]. Refinement 2: [refinement 2].' \\
After that, assist the user in a regular chat fashion without following a specific response format.\\
The complaint you are addressing is: `[...]'
\end{tcolorbox}

To assess whether utterances referred to the AI assistant directly, we used a Language Model as Research Assistant (LMRA) \cite{eloundou2024first,riedl2025quantifying} with the prompt below.
\begin{tcolorbox}[llmprompt, title=LMRA Assessment ``is this utterance about AI'' (gpt-5-mini)]
You are an expert research assistant. Your task is to evaluate the utterance below and assess whether it contains an explicit reference to an AI entity. Answer only yes or no.\\
Utterance: [...]
\end{tcolorbox}

\paragraph{Analysis Details.} Table \ref{tab:words} shows most influential AI exposure words words (by count differences) across the two treatments.
\begin{table}[h!]
    \begin{tabular}{@{\extracolsep{5pt}}l l l l l l }
    \toprule
    \multicolumn{4}{c}{Most Telling Words for} & \multicolumn{2}{c}{Most Common \&} \\ 
    \multicolumn{2}{c}{Formal Bot} & \multicolumn{2}{c}{Empathetic Bot} & \multicolumn{2}{c}{Least Differentiating} \\ 
    \cline{1-2} \cline{3-4} \cline{5-6}
    regarding         & (32 vs.~3) & sorry            & (3 vs.~35) & know       & (32 vs.~36) \\
    formal            & (25 vs.~0) & truly            & (0 vs.~32) & let        & (32 vs.~35) \\
    acknowledge       & (26 vs.~2) & frustrating      & (1 vs.~28) & like       & (33 vs.~34) \\
    communication     & (28 vs.~4) & completely       & (0 vs.~26) & response   & (34 vs.~30) \\ 
    report            & (22 vs.~0) & hear             & (2 vs.~26) & issue      & (33 vs.~29) \\
    investigation     & (23 vs.~2) & right            & (0 vs.~24) & please     & (32 vs.~30) \\
    company           & (23 vs.~3) & make             & (4 vs.~26) & anything   & (29 vs.~31) \\
    professional      & (20 vs.~0) & 've (e.g., I've) & (3 vs.~25) & help       & (24 vs.~35) \\
    currently         & (19 vs.~1) & get              & (2 vs.~23) & need       & (33 vs.~26) \\
    team              & (31 vs.~13)& understand       & (12 vs.~30)& assistance & (32 vs.~24) \\
    \bottomrule
    \end{tabular}
    \caption{Word frequency contrast (count difference) of AI exposure words in each treatment condition. Parentheses provide counts in Formal vs.~Empathetic Bot conditions.}
    \label{tab:words}
\end{table}

\begin{figure}[h!]
  \centering
  \includegraphics[width=.49\textwidth]{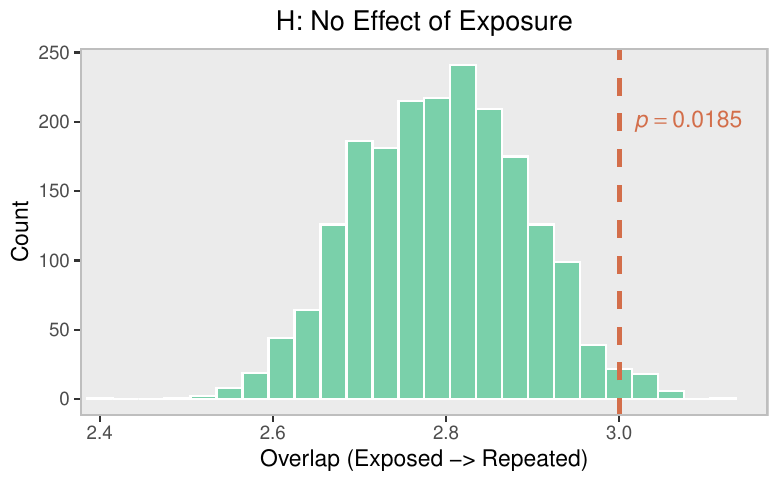}
  \includegraphics[width=.49\textwidth]{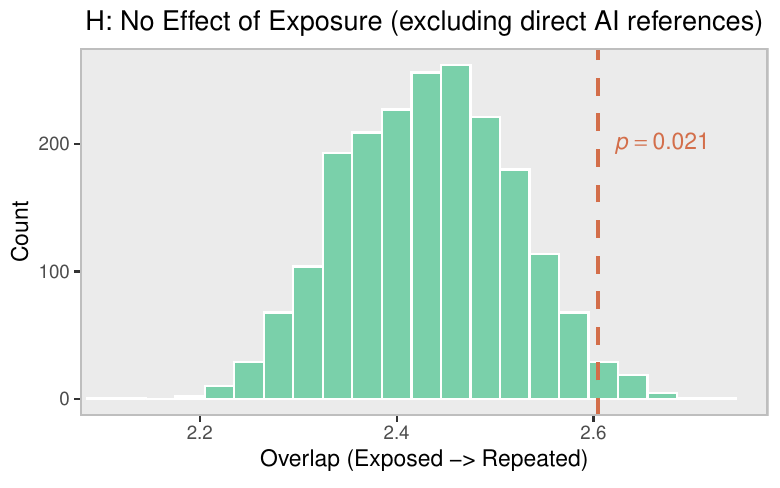}
  \caption{Permutation test assessing observed reuse of AI introduced words in human-human conversation against distribution of permutation null values in which treatment and counterfactual exposures are randomly shuffled.  Bars show null distribution, red dashed line shows observed value.
  \textbf{Left:} Results using full data.
  \textbf{Right:} Subset after removing utterances that directly reference AI.
  ($B=2,000$ permutation samples; one-sided $p$-value).}
  \label{fig:permutation}
  \Description{Permutation test assessing observed reuse of AI introduced words in human-human conversation against distribution of permutation null values in which treatment and counterfactual exposures are randomly shuffled.}
\end{figure} 

For the Poisson-based permutation analysis we used the following regression equation:


\begin{equation}
\label{eq:study2regresion}
\begin{aligned}
Y_{u} \mid \mathbf{x}_{u}, \alpha_{i(u)}, \delta_{u(u)} &\sim \operatorname{Poisson}(\mu_{u}),\\
\log \mu_{u} &= \beta_0
+ \beta_1\,\mathrm{UtteranceWordsLog}_{u}
+ \beta_2\,\mathrm{Type}_{u}
+ \alpha_{i(u)} + \delta_{u(u)} \, 
\end{aligned}
\end{equation}

where $u$ indexes utterances, $i(u)$ is the individual $i$ who spoke utterance $u$, Type is a dummy variable indicating whether the instance is observed vs.~counterfactual. We include fixed effects for the individual and the utterance identifier (first, second, ... etc ), and cluster-robust (sandwich) standard errors clustered on the individual-utterance level to account for dependency in the paired observed-counterfactual construction.

To complement analyses based on the debrief interview, we also included a written pre- vs.~post-test in the AI interaction phase (this analysis is mentioned in the pre-registration). Specifically, before interacting with the AI assistant and before seeing any of the customer complaints they would work on, subjects answered the following survey questions with freeform text:
\textit{Imagine you're part of a customer service team. What strategies do you think your team should adopt when handling complaints?} Note again that the questions use team-level ``we'' framing to index collective constructs \citep{mathieu2022indexing}. After the AI interaction, subjects were asked the question a second time. We take each user's written responses, compute text embeddings using OpenAI's \verb|text-embedding-3-large| model, then perform a permutation test comparing cosine similarity between post-treatment responses written by users assigned to the same treatment condition versus a null distribution in which treatment assignment is randomly shuffled. We find no significant difference ($p=0.442$). This analysis has low power to detect effects given that responses are quite short and there is large between-individual variation. To gain more leverage in the analysis we perform additional post-hoc analyses using the directional embedding shift between user's pre-exposure response and their post-exposure response. We then perform Two-sample Hotelling's T-squared test \citep{hotelling1931generalization} to see if the distributions are different across the two treatment conditions (i.e., whether the treatments induce different directional shift). Since embeddings have 3072 dimensions and the Hotelling test is limited to dimensionality of the number of observations (3072 is much larger than the number of observations in our study) we need to reduce the embedding dimensionality. We perform PCA with $k=1, 2, \ldots, 40$ components. The smallest $p$-value is $0.0428$ ($k=33)$, 10 $p$-values are marginally significant at 10\% significance level, correcting for multiple testing with Cauchy combination we find $p=0.142$, and an uncorrected Fisher test is highly significant at $p = 4.2 \times 10^{-5}$.


\section{Study 2 - Additional Study Details and Results} \label{appendix:study1}

\subsection{Experimental Design and Procedure}
We recruited subjects through fliers, email lists, and word of mouth from the Northeast United States to participate in this study. Team size varied between three and four people based on convenience: While we scheduled four participants for each session, we proceeded if at least three were present. Subjects were predominantly young, with a mean age of 21.7 years, and skewed female.

Participants first received written instructions and a short practice trial to familiarize themselves with the interface and task structure. Teams were told that an AI assistant, described as a ``Puzzle Master’s Helper,'' would provide occasional hints. They were not informed about the helpfulness manipulation. After the task, participants completed a survey on trust, perceptions of team membership, and shared mental models.

Teams had 40 minutes to solve the entire puzzle. The pattern recognition task is designed to be challenging; most teams fail to reach all milestones. To help them, we introduce an AI assistant called the ``AI Puzzle Master'' that dispenses one clue for each milestone. The AI assistant was not interactive but instead delivered pre-recorded clues at fixed time intervals. The AI assistant was a disembodied entity, represented by a black screen displaying its name in the online video conferencing software. We created four different AI assistant recordings, one for each treatment condition, to ensure uniformity in message content, sequence, and timing (listen to recordings \href{https://osf.io/4ywp5/?view_only=f0440192455046ea9d071f357c44d415}{here}). Each statement delivered by the AI assistant is an intervention driving the causal identification of effects in our statistical analysis. 

We manipulated the content of the AI assistant messages (helpful or unhelpful) and the AI assistant's voice (human or robotic sounding). Each clue was designed to encourage teams to explore patterns among the symbols associated with each treasure chest. The helpful AI assistant provided relevant information that could help teams uncover milestone patterns (e.g., ``the octopus rule relates to separation between the gems''). These helpful clues only pointed teams in the right direction, they did not directly reveal the milestone answer. In the unhelpful condition, the AI assistant provided information that was task-related, not helpful to uncovering the correct milestone pattern and included instances where the AI provided incorrect information (e.g., suggesting that the ``octopus rule'' has to do with the sizes of the gems, when gem size is, in fact, not important). For the human voice condition, we hired a female voice actor to record the audio. For the robotic voice, we used a text-to-speech program which had a female timbre but lacked human prosody and paralanguage.

Each clue referenced one of the puzzle’s milestones (e.g., ``the octopus curse relates to the sizes of the gems''). Clues varied in accuracy based on the assigned helpfulness condition. The assistant did not respond to team input, ensuring that its influence could be analyzed in terms of linguistic priming rather than real-time responsiveness. The AI assistant interjecting at pre-determined time points (approximately every 5–7 minutes). The task is particularly suitable for studying the development of shared language because it presents participants with several unfamiliar symbols for which they do not have established terminology. Since the symbols can be described using a variety of words, teams must develop a shared vocabulary to effectively distinguish them in order to direct each other's attention and communicate precisely \cite{clark1986referring,reagans2023shared}. Several other studies investigating the development of shared language have relied on similar symbol tasks \cite{clark1986referring,reagans2023shared}. 

In addition to the arguments provided in the main manuscript, there are several additional reasons to justify that this is a suitable and effective design to study the effect of an AI assistant on collective attention of human teams. 
First, the focus of our study is human-human communication. Whereas our agent does not meet the definition of agency required for most definitions of human-agent teams \cite{mcneese2018teaming} this trade-off allows us to precisely attribute changes in teams' cognitive alignment to the AI assistant's interventions. If the agent was autonomous, the endogenous co-creation of cognitive alignment would substantially impede our ability to (causally) attribute emergent alignment to the AI treatment intervention. By controlling the timing and content of the AI assistant's interjections, we limit the confounding effects of endogenous processes that would arise if the AI possessed agency \cite[e.g.,][]{schelble2022let,zhang2021ideal}. It also reduces variability in interaction patterns and sharpens focus on human behavior. This decision enhances the internal validity of our study, at the expense of external generalizability. 

Second, current AI systems still have limited autonomy capabilities and it may take more time before they qualify as real team members \cite[e.g., in the sense as defined by][]{klien2004ten}. Using designs with limited agency can help avoiding over-reliance on technology readiness and lower technical hurdles of studying phenomena such as human-AI interaction in team settings \cite{schecter2023vero}. 
Third, even systems with limited autonomy can significantly affect feedback dynamics in human-technology systems and allow us to study how users modify their behavior when interacting with AI \cite{klien2004ten}. In fact, often trust in simpler, deterministic agents is often higher than in more complex but more opaque and less predictable agents \cite{klien2004ten}. When systems takes the initiative to adapt to its users' working styles, users have been shown to adapt their own behavior less because of confusions that the interaction creates \cite{klien2004ten}. While the AI agent's interjections may seem deterministic, this design transparently assigns the AI assistant unique role on the team and makes it clear what capability and limitations the assistant has. This is in line with the task framing in which the puzzle master would ``reveal clues to you as you progress through the challenge. You do not have the option of requesting clues, only receiving.'' This facilitates participants' coordination among each other and with AI teammates, especially at the beginning of the task. 

Third, predefined interaction scripts this design is consistent with similar designs used to study human-AI teaming designs \cite[e.g.,][]{zhang2023investigating,schecter2023vero}. 
Fourth, our design mimics recent developments with in which groups of humans are supported by a centralized AI agent such as adding virtual assistants to Microsoft Teams group chats (e.g., to send automated meeting summaries or reminders)\footnote{\url{https://learn.microsoft.com/en-us/microsoftteams/platform/samples/virtual-assistant}} or AI assistants added to Slack channels.\footnote{\url{https://runbear.io/solutions/use-cases/internal-support?utm_source=astro}} In such settings, the decision of the AI agent to speak up can intentionally or unintentionally shift a group's focus of attention and may subtly affect alignment on shared language.

\subsection{Processing of Audio Data}
\begin{figure}[h]
  \centering
  \includegraphics[width=1\linewidth]{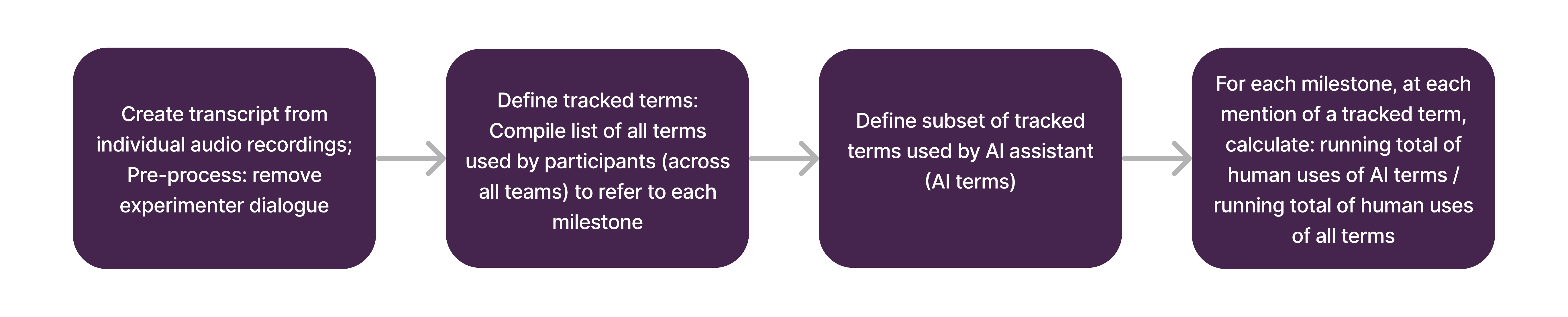}
  \caption{Steps of data pre-processing and variable construction.}
  \label{fig:DataProcessing}
  \Description{Steps of data pre-processing and variable construction.}
\end{figure}

To measure the AI assistant's impact on team collective attention and linguistic alignment, we transcribed the audio recordings of team communications and identified (using a single coder, using inductive coding, using a predetermined list set of eight objects (six curses plus two peripheral referents ``gem'' and ``symbol'') each instance of a team member referring to one of the task milestones (Fig.~\ref{fig:DataProcessing}). We compiled a complete list of terms used by participants to refer to each of the milestones, the symbols, and the gems, including different terms used to refer to the same thing. This list of terms formed our set of \textit{tracked terms}. Each time a participant used one of the tracked terms to refer to a milestone, a symbol, or a gem, we recorded its use, forming the basis of our lexical alignment measure (see next section).

The AI assistant consistently used the same language to refer to each of the milestones and to the gems. This allows us to analyze the team’s term usage to see whether they used the same terminology as the AI or different terminology. Every time we updated the running usage counter for one of the terms, we also updated a running lexical alignment score based on whether or not the term matched the AI’s term for that referent. The alignment score increased if the term was an AI term and decreased if it was not (unless it was already at 100\% or 0\%, in which case it did not change).

Among the tracked referents, we distinguish between five core referents and two peripheral referents. The five referents correspond to the five curses, each of which is the subject of one clue from the AI assistant. Since the core referents are each directly related to uncovering the pattern among symbols, the AI assistant's interjection can be identified as high or~low quality depending on the helpful vs.~unhelpful treatment condition. The peripheral referents are ``symbols'' and ``gems'', which are relevant to the task but are not milestones themselves. Gems are mentioned in several AI interjections, while symbols are never mentioned, making the symbol referent an appropriate baseline from which changes in collective attention can be contrasted. 

In summary, each referent corresponds to a set of tracked terms and each referent is mentioned by the AI assistant in one of its interjections. Following the interjection, teams may talk more (or less) about the mentioned referent, and they may do so using either the same term used by the AI or use a different term to refer to the referent. For example, while the AI assistant used the word ``duel'' a team may instead refer to the symbol as ``swords''.

\subsection{Analysis Method}
Human communication is a dynamic process. A key challenge in our analysis is to \textit{causally} attribute specific changes in this fluid team communication to the interventions of the AI assistant. To achieve this causal identification, we employ a panel regression using observations on the utterance level following a difference-in-difference (DD) framework \cite{angrist2009mostly}. Difference-in-difference is a standard econometric approach for causal identification in panel data settings, i.e., settings with repeat observations \cite{angrist2009mostly}. 
Specifically, we consider term use (as captured by the lexical alignment measure) in segments right before the AI intervention and right after for each of the six AI intervention (e.g., Segment 1 surrounding the ``computer hex'' clue; see bottom of Fig.~\ref{fig:team7}). The DD analysis then compares the before/after change in AI term usage during a specific segment to the same before/after change during the same segment of a term that was \textit{not} used by the AI. For example, the before/after change in the usage of the term ``octopus'' during the segment surrounding the ``pirate clue'' serves as a no-treatment control 
 (because the ``pirate clue'' did not mention ``octopus'') 
to the before/after change in term usage of ``pirate'' 
 (which could have been affected by the AI's mention of ``pirate''). 
Formally, we estimate the following regression equation 

\begin{align}
\begin{split}
    \textit{AI Alignment}_{ogs} = & \beta_1\textit{After}_{ogs} \\
    + & \beta_2\textit{After}_{ogs}\times \textit{Helpful}_{g} \\
    + & \beta_3\textit{After}_{ogs}\times \textit{Robotic}_{g} \\
    + & \beta_4 \textit{ObjectMentions}_{og} + \beta_5 \textit{ElapsedTime}_{og} \\ 
    + & \alpha_o + \alpha_g + \alpha_s + \epsilon_{ogs}
    \label{eq:milestone_regression}
\end{split}
\end{align}

\noindent where \textit{After} is a dummy variable indicating whether the utterance related to object $o$ happened before or after the AI intervention related to object $o$, \textit{Helpful} is a dummy indicator whether the team $g$ was assigned to the helpful (vs.~unhelpful) AI condition condition, \textit{Robotic} is a dummy indicator whether the team $g$ was assigned to the human-sounding (vs.~robotic sounding) voice condition condition, $\alpha_o$ are object-level fixed effects,  $\alpha_g$ are group-level fixed effects, $\alpha_s$ are segment-level fixed effects, and $\epsilon_{ogs}$ are error terms. We clustered standard errors on the team-level to handling within-cluster correlation of repeated observations.
The coefficient $\beta_1\textit{After}$ captures the causal effect of the AI intervention. Notice the main effect of the treatment will drop out due to the team-level fixed effect and only the interaction term remains. The the coefficient $\beta_2$ for the interaction term of $\textit{After} \times \textit{Helpful}$ captures the differential effect of the helpful treatment condition (compared to the unhelpful condition), and  $\beta_3$ for the interaction term of $\textit{After} \times \textit{Robotic}$ captures the differential effect of the robotic-sounding voice condition (compared to human-sounding). Estimates are driven by six before/after comparisons created by the six AI interventions, each one of them comparing five unaffected terms serves as control contrasted from one term affected by the intervention.
We include the count of \textit{ObjectMentions} and the \textit{ElapsedTime} (in minutes) as control variables. 
We estimate variations of this equation where we substitute \textit{Humanness} and \textit{Treatment} as key variables of interest for \textit{Helpful}. We report standard errors clustered at the team-level.

\subsection{Results}

\subsubsection{Manipulation Validation} 
To confirm that the robotic voice sounded non-human, we recruited 400 workers from Amazon Mechanical Turk for a validation test. We used eight different labels to describe the robotic voice: a machine, a computer, a robot, a digital assistant, an automated assistant, automation, and artificial intelligence. Each label was tested by 50 workers. Workers listened to audio clips of the robotic voice's clues and rated on a seven-point Likert scale how human or non-human the voice sounded. Each group answered a question based on their assigned label (e.g., ``Is this voice a machine or a human?''). The results showed that workers consistently rated the robotic voice as non-human, no matter which label was used. For example, at the less significant high end, the label ``robot" had results of $z = 2.263$, $p = 0.024$, while at the more significant low end, the label ``computer" had results of $z = 3.677$, $p < 0.001$.

\subsubsection{Example Case Studies}
We start our analysis with an example from Team 18, a team of three people (plus the AI) in the \textit{Human-Unhelpful} condition. After receiving their instructions, the team begins the task by looking for patterns among the gems, referring to them as ``diamonds.'' The team points out various features of the treasure chests and notices some similarities between chests with the same curse. During the first fifteen minutes, they refer to the gems seven times, always using the word ``diamond.''  Nobody uses the word ``gem'' until the AI assistant gives its first clue at 15:07. In this case, the clue is unhelpful and about the computer hex milestone, but it does include a mention of ``gems.''

After the AI's interjection, the team's focus shifts to discussing the computer hex curse, which they had not talked about before. This shows how the AI affects the timing of \textit{what} teams talk about, addressing RQ1.  The team expresses confusion because they cannot see the pattern suggested by the AI assistant. Despite this, they stop using the word ``diamond'' and start using the word ``gem.'' Between the first and second AI interjections, there are eight more mentions of the gems, and all of them use the term ``gem.'' This shows that the team has adopted the AI's language for the gem referent, addressing RQ2 about how the AI influences \textit{how} teams talk.

At 20:07, the AI gives its second clue, stating that the octopus milestone is related to the sizes of the gems. However, this clue contradicts the pattern the team has started to identify, causing them to become suspicious of the AI assistant.  One participant suggests that the clues might be designed to mislead them, but another team member responds by saying, ``We have to trust the Puzzle Master.''

At 30:07, the AI gives its fourth clue, which is about the duel milestone. After hearing the clue, a team member again suggests that the AI's clues might not be helpful. Throughout the rest of the task, the team switches between using the terms ``duel'' and ``sword'' to talk about the duel clue. They do not fully adopt the term ``duel'' suggested by the AI assistant for this core referent. In the end, the team is unable to complete the task within the given time.

A different team, Team 19, assigned to the \textit{Robotic-Helpful} condition, shows a different pattern. This team demonstrated higher trust in their AI assistant, which gave them helpful information instead of unhelpful clues.  

The first two clues, about the ``computer hex'' and ``octopus'' milestones, used terms that matched the team's own language at the start of the experiment. Because of this, the team did not change the terms they were using. Both clues matched the group's observations, so they likely had no reason to doubt the AI's reliability.  

The third clue, given at 24:34, informed the team that the ``quadruple bypass'' milestone related to the colors of the gems. At this point, a team member explained that the ``quadruple bypass'' milestone was the same as what the team had been calling ``heartbreak'' or ``heart.'' Despite trusting the AI, the team continued to use their own terms, ``heartbreak'' or ``heart,'' instead of switching to ``quadruple bypass.''  Similarly, the team kept using ``pirate'' or ``pirate eye'' when discussing the ``eye patch'' milestone, even though the AI gave a helpful clue using the term ``eye patch.''

\subsubsection{Shared Mental Models}
We collect data on several constructs through a post-experiment survey. We measure the alignment of mental models as the degree to which members' mental models are consistent \cite{mohammed2010metaphor} using six items from \cite{johnson2007measuring}. Items propose statements like ``my team knows specific strategies for completing various tasks'' and participants are asked to respond on a Likert scale. With this data, we compute an intragroup agreement index within each item using standard deviation (to measure how consistently within-team responses overlap versus diverge), and then average the standard deviations across the six items. 

\begin{figure}
    \centering
    \includegraphics[width=.35\linewidth]{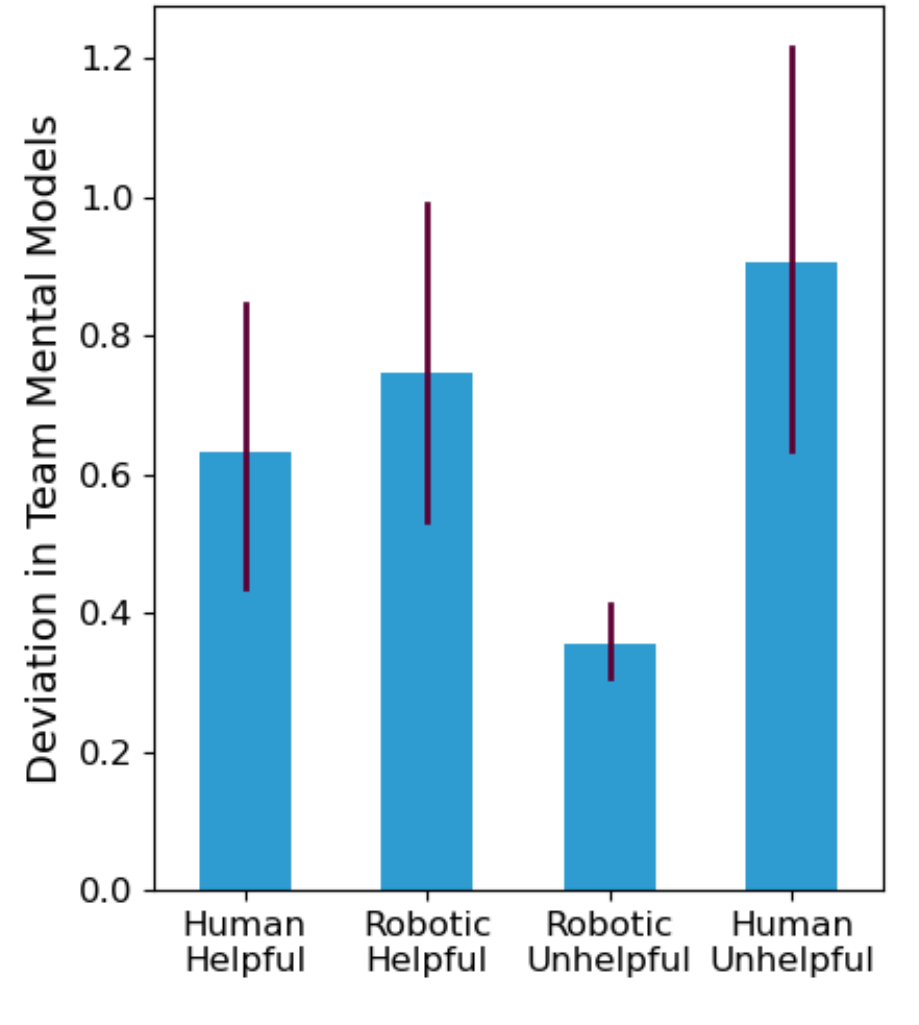}
    \caption{Alignment of mental models across treatments. Bars show strength of mental model alignment (measured as standard deviation across answers to shared mental model questions per treatment; lower values mean more consistent intragroup answers indicating stronger alignment; higher values indicate lower alignment). Strong alignment in the \textit{Robotic-Unhelpful} condition points to exclusion of AI from the team.}
    \Description{Bar graph of shared mental model alignment.}
    \label{fig:SMM}
\end{figure}

\subsubsection{Additional Explanations for Observed Cognitive Alignment}
To explore possible explanations for the cognitive alignment effects we observe, we investigate two predictor variables: trust in AI and perception of AI assistant as a teammate.

\paragraph{Trust in AI} \label{sec:trust}
We measure trust in the AI assistant and perceptions of the assistant's contribution quality, we use six items adapted from \cite{pan2017impact} (originally designed to measure trust in online avatars). Participants responded on a five-point Likert scale from Strongly Disagree to Strongly Agree to items like ``The Puzzle Master gave helpful clues.'' Subjects rated the unhelpful AI as significantly less trustworthy ($p < 0.001$) and less intelligent ($p < 0.001$) than the helpful AI. Voice anthropomorphism did not have a significant impact on participant impressions of trustworthiness ($p = 0.570$), but the ratings were slightly lower in the Robotic treatments.
The greater level of trust in the helpful AI conditions may explain why participants aligned more with the AI assistant's language for core referents in these conditions. Humans are more likely to adopt recommendations by trustworthy agents that provide high-quality explanations \cite{wang2007recommendation}. Overall, a stark pattern emerges: even in the unhelpful conditions in which trust assessments of the AI assistant were significantly lower, the AI assistant directed team attention to the specific task aspects it mentioned (see Section \ref{sec:colAtt}) and influenced their lexical choices.

\paragraph{Is the AI Assistant a Part of the Team?}  \label{sec:partofteam}
The post-experiment survey asked subjects to report whether they felt that the AI assistant was part of their team or even a leader of their team (answered using a single Likert scale item). Humans did not consider the AI to be a part of their team, with 77\% of participants responding ``Strongly Disagree'' or ``Somewhat Disagree.'' Humans responded even more negatively when asked whether the AI was a leader of their team, with 93\% of participants responding ``Strongly Disagree'' or ``Somewhat Disagree.'' Ratings were especially low in the \textit{Robotic-Unhelpful} condition, where 96\% did not consider the AI assistant to be a leader of the team and 93\% of participants did not consider it to be part of the team at all. This suggests that the AI assistant's substantial influence over teams' collective attention and lexical alignment is not predicated on its inclusion as a team member, but rather occurs despite it being considered an outsider. This phenomenon indicates a path through which AI assistants can---intentionally or unintentionally---affect intra-team communication and cognitive alignment. By presenting as a non-human entity, AI assistants may strengthen the bond among human team members (the in-group) by serving as a contrasting out-group, giving rise to an ``us vs.~them'' dynamic \cite{gallotti2013social,babel2010dialect}. The degree to which AI assistants are (or can be) excluded from the team may further shape cognitive alignment among human teammates and enhance joint action. The next section provides additional evidence for this exclusionary dynamic through an analysis of pronoun use.

\end{document}